\documentclass[twocolumn,superscriptaddress, prl]{revtex4-1}
\usepackage[T1]{fontenc}
\usepackage[latin9]{inputenc}
\usepackage{color}
\usepackage{units}
\usepackage{mathrsfs}
\usepackage{amsmath}
\usepackage{amssymb}
\usepackage{graphicx}
\usepackage[english]{babel} 

\makeatletter
\usepackage[caption=false]{subfig}

\@ifundefined{showcaptionsetup}{}{%
	\PassOptionsToPackage{caption=false}{subfig}}
\usepackage{subfig}
\makeatother

\usepackage{babel} 

\begin{document}
\title{Scattering-based geometric shaping of photon-photon interactions }
\author{Shahaf Asban}
\email{Shahaf.S.Asban@gmail.com}

\affiliation{Department of Chemistry and Physics and Astronomy, University of California,
Irvine, California 92697-2025, USA}
\author{Shaul Mukamel}
\email{smukamel@uci.edu}

\affiliation{Department of Chemistry and Physics and Astronomy, University of California,
Irvine, California 92697-2025, USA}
\begin{abstract}
We construct an effective Hamiltonian of interacting bosons, based on scattered
radiation off vibrational modes of designed molecular architectures.
Making use of the infinite yet countable set of spatial modes representing the scattering of light, we obtain a variable photon-photon interaction in this basis. The effective Hamiltonian hermiticity is controlled by a geometric factor set by the overlaps of spatial modes. Using this mapping, we relate intensity measurements of the light to correlation functions of the interacting bosons evolving according to the effective Hamiltonian, rendering local as well as nonlocal observables accessible. This architecture may be used to simulate the dynamics of interacting bosons, as well as designing tool for multi-qubit photonic gates in quantum computing applications. Variable hopping, interaction and confinement of the active space of the bosons are demonstrated on a model system.  

\end{abstract}
\maketitle

Quantum machines  are fundamentally different from their classical
counterparts \citep{nielsen_chuang_2010}. Classical computers are
implemented using a binary basis set. One of the benefits of this
choice is minimal average energy consumption \footnote{Assuming voltage levels are assigned to bit states, having '0' represented by zero voltage.} as well as minimal bit error rate in the information transmission 
of a noisy channel \citep{massoud2007digital}. For such a machine  
to be useful, it requires large scale integration of fundamental operations,
where each level adds to the overall error rate. Deterministic intermediate
quantities can be measured and corrected via feedback loops without
interrupting the calculation process. In quantum machines the smallest
possible unit of data (qubit) carries a phase which manifests a continuous
degree of freedom. Upscaling of operations on qubits is also required
for nontrivial tasks. The propagation of quantum information through
such integrated system evolves errors continuously as well. This makes
the realization of fault tolerant quantum processing challenging,
and continues to motivate intense scientific effort \citep{PhysRevA.52.R2493,PhysRevA.64.012310,Schindler1059}.
Quantum simulators based on optical traps pioneered by Cirac and
Zoller \citep{PhysRevLett.74.4091,PhysRevLett.81.3108} have matured
experimentally \citep{Monroe2002,Kim2010}. They are widely used in
the study of quantum dynamics such as spin frustration \citep{Kim2010}
and thermalization and localization transitions \citep{Schreiber842,Smith2016}.
Recently more applications of lattice gauge theories have been proposed,
appealing to simulations of high energy physics \citep{Zohar_2015,RICO2018,Muschik_2017}.
While these fascinating quantum simulators offer unprecedented glimpse
into nonequilibrium dynamics, upscaling the number of qubits is just
as challenging. 

\begin{figure}[th]
\includegraphics[bb=70bp 330bp 550bp 520bp,clip,scale=0.54]{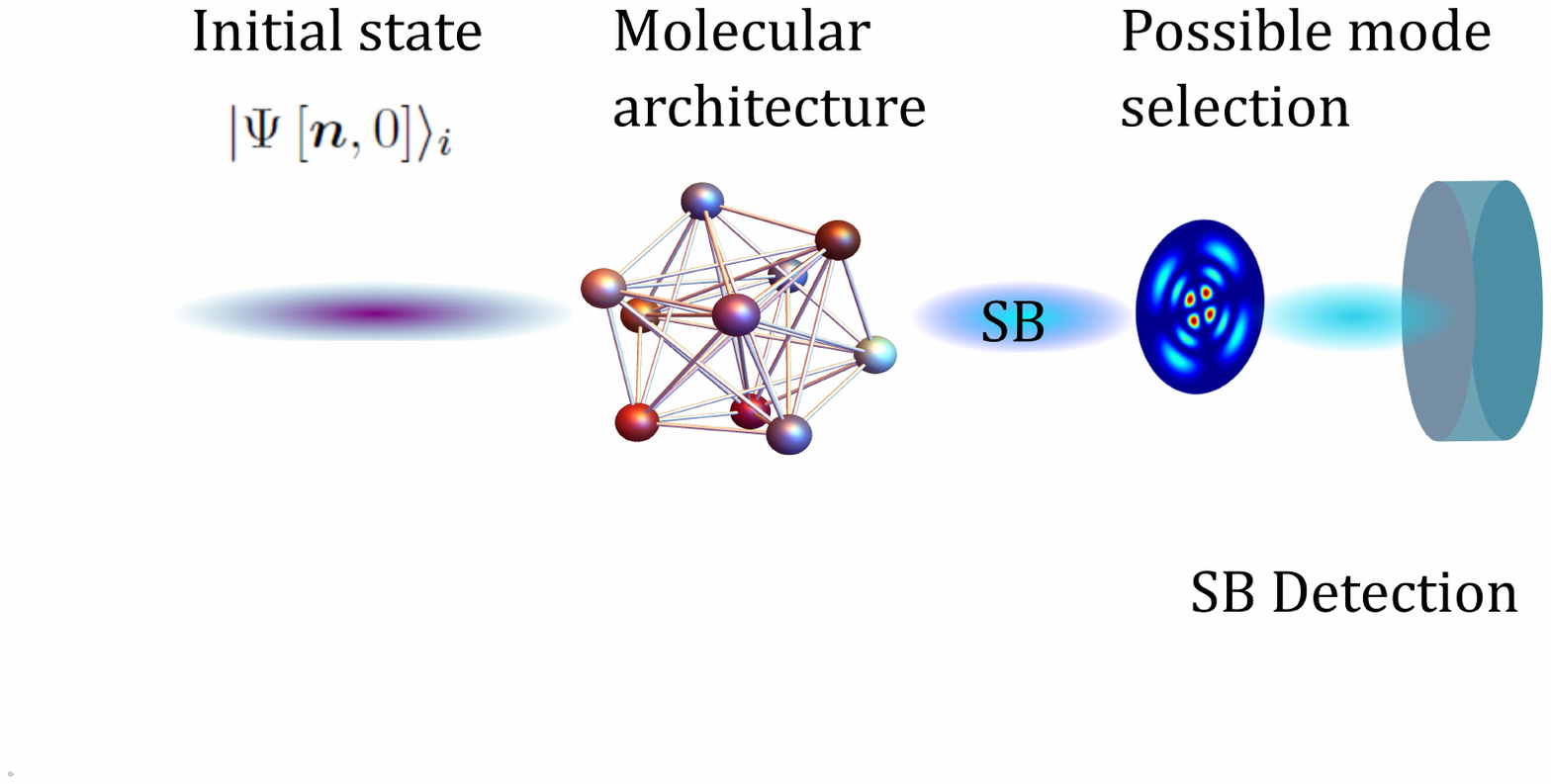}

\caption{\emph{Proposed realization of the geometric SPC}. The incident beam is scattered-off
a molecular architecture, realizing interaction between its 
spatial components denoted SB. A single mode as well as any combination of such are addressable, rendering projective measurement in the chosen basis possible. The SBs can be mapped into discrete lattice sites for which cross-correlations can be constructed using the photon-statistics. \label{fig:Setup}}
\end{figure}

Here we propose a Geometric scattering-based Spatial Photon Coupler (SPC) that can be used to simulate quantum dynamics of interacting bosons. The setup is depicted
in Fig.$\left(\text{\ref{fig:Setup}}\right)$, and based on \emph{off-resonant
scattering} of photons on geometrically arranged
distribution of molecules. We show that the scattering process can
be mapped into the effective Hamiltonian,

\begin{equation}
H_{eff}=\sum_{nk}\Theta_{nk}a_{n}^{\dagger}a_{k}-\sum_{nklm}U_{nklm}a_{n}^{\dagger}a_{k}a_{l}^{\dagger}a_{m},\label{Effective Hamiltoian}
\end{equation}

\noindent where $a_{k}\left(a_{k}^{\dagger}\right)$ is a discrete
bosonic annihilation (creation) operator. The hopping $\left(\Theta_{nk}\right)$
and the interaction $\left(U_{nklm}\right)$ terms are determined
by three main quantities: (1) the geometry (design) of the microscopic
building blocks (2) their internal structure (spectrum) (3) the measurement
basis chosen. It holds two significant advantages. First, it can be designed to maintain hermiticity such that marginal losses in photon number occur \footnote{In linear order of the paraxiality parameter $\vartheta$ there are \emph{no losses} in the longitudinal modes. Losses in the transverse modes are governed solely by the geometry. Higher orders exhibit losses and demonstrated in the illustration below.  }, inelastic contributions in this case result in frequency shift of the incident photon. Second, while the number of connected modes is infinite, it can be confined with intelligent design. Our goal is to shape the induced dynamics constrained by the Hamiltonian in 
Eq.$\left(\text{\ref{Effective Hamiltoian}}\right)$ then read the encoded information from the final photonic wavefunction,

\begin{equation}
\vert\Psi\left[\boldsymbol{n},\tau\right]\rangle_{f}=e^{-iH_{eff}\tau}\vert\Psi\left[\boldsymbol{n},0\right]\rangle_{i}.\label{Target state}
\end{equation}

\noindent Intensity measurements  reveal the Scattered Bosons (SBs) densities $\hat{n}_{k}\equiv a_{k}^{\dagger}a_{k}$,
evolving on a network (graph) of a topology imposed by $H_{eff}$
connectivity as depicted in Fig.$\left(\text{\ref{fig:Setup}}\right)$.
$\tau$ denotes the interaction time, beyond which $U_{nklm}\equiv0$.

We consider arbitrary initial superposition of spatial modes, reflecting some initial (multiple-photon) distribution of SBs. One possibility of particular interest to the simulation of interacting bosons, is a (normalized) product of $N$ single photon states $\vert\boldsymbol{1}\rangle=\sum_{n}{\cal C}_{n}a_{n}^{\dagger}\vert0\rangle$, corresponding to initially noninteracting bosons. One way to achieve that is by direct state preparation. Another is via two-mode spontaneous parametric down converter source in which one photon is scattered while the other serves as the Reference for the SB (RSB) as was done in \cite{Asban2019} and inspired by  \citep{Fickler2016,Fickler640,PhysRevA.87.012326,Bavaresco2018,Straupe2011}, see \footnote{Using entangled photon pair such that the as SB and RSB ideally results in background free signal yet harder to produce. }.
In both manifestations, the RSBs are used to extract the SBs statistics which is first scattered as depicted in Fig $\left(\text{\ref{fig:Setup}}\right)$.
Surveying the SBs statistics using cross-correlations with RSBs, renders
"local" $\propto\left\langle \hat{n}_{r}\left(\tau\right)\right\rangle $ and
 $\left\langle \hat{n}_{r}\left(0\right)\hat{n}_{r}\left(\tau\right)\right\rangle $, as well as "nonlocal"
quantities $\propto\sum_{R}\left\langle \hat{n}_{r}\left(0\right)\hat{n}_{r+R}\left(\tau\right)\right\rangle $ accessible. While conventional quantum simulators are limited by the
number of physical qubits, the proposed geometric SPC benefits from controlled
- potentially infinite - number of participating modes. These characteristics
are highly desirable for simulating thermodynamic properties of
interacting particles. 

\emph{Constructing the effective Hamiltonian.} Off-resonant light-matter interaction
is given by the minimal coupling Hamiltonian,

\begin{equation}
H_{\mu\phi}=\int d\boldsymbol{r}\sigma\left(\boldsymbol{r},t\right)\boldsymbol{A}^{2}\left(\boldsymbol{r},t\right),
\end{equation}

\noindent where $\mu$ and $\phi$ denote the matter and photon
fields respectively. Radiation modes interaction is mediated by the
charge-density operator $\sigma\left(\boldsymbol{r}\right)$. The
vector potential in the paraxial approximation takes the form $\boldsymbol{A}_{p}\left(\boldsymbol{r},t\right)=\sum_{\sigma,l,p}\underset{0}{\overset{\infty}{\int}}dk_{0}C\left(k_{0}\right)\left[A_{\sigma lp}^{\left(+\right)}\left(k_{0}\right)e^{ik_{0}\left(z-ct\right)}+h.c.\right],$
where $k_{0}$ is the wavector in the longitudinal direction, $C\left(k_{0}\right)=\left(\nicefrac{1+\vartheta^{2}}{16\pi^{3}\varepsilon_{0}k_{0}}\right)^{\nicefrac{1}{2}}$
and $\vartheta=\nicefrac{q}{\sqrt{2k_{0}^{2}}}$ is the degree of
paraxiality \citep{PhysRevA.72.060101,Calvo2006}. The vector potential
operator is defined by $A_{\sigma lp}^{\left(+\right)}\left(k_{0}\right)=\epsilon_{\sigma}\hat{a}_{\sigma,l,p}\left(k_{0}\right)\psi_{l,p}\left(\boldsymbol{\rho},z;k_{0}\right)$
where $\hbar=c=1$. $\boldsymbol{\rho}$ is the polar distance (cylindrical
coordinates) and the indices $l,p$ label the spatial basis set $\psi_{lp}\left(\boldsymbol{\rho},z;k_{0}\right)$
{[}in \citep{PhysRevA.72.060101,Calvo2006} Laguerre-Gauss (LG) basis
is used{]}. Any complete basis which is a solution of the paraxial
equation, such as Hermite- or Ince-Gauss (HG,IG) can be used. Here
$\epsilon_{\sigma}=\nicefrac{\left(\boldsymbol{u}_{x}-i\sigma\boldsymbol{u_{y}}\right)}{\sqrt{2}}$
is the polarization vector {[}circular for LG modes (linear for HG){]}
and $\sigma=\pm1$. The field (creation/annihilation) operators in
the paraxial basis satisfy the canonical commutation relations, $\left[\hat{a}_{\sigma,l,p}\left(k_{0}\right),\hat{a}_{\sigma',l',p'}^{\dagger}\left(k_{0}^{'}\right)\right]=\delta_{\sigma\sigma'}\delta_{ll'}\delta_{pp'}\delta\left(k_{0}-k_{0}^{'}\right)$.
The paraxial basis is given by the transformation of the standard
operators $\hat{a}_{\sigma,l,p}\left(k_{0}\right)=\int d^{2}\boldsymbol{q}\phi_{lp}^{*}\left(\boldsymbol{q}\right)\hat{a}_{\sigma}\left(\boldsymbol{q}\right).$
This basis has the following properties,

\begin{subequations}

\noindent 
\begin{eqnarray}
\sum_{l,p}\phi_{lp}^{*}\left(\boldsymbol{q}\right)\psi_{lp}\left(\boldsymbol{\rho},z,k_{0}\right) & = & e^{i\boldsymbol{q}\cdot\boldsymbol{\rho}-k_{0}\vartheta^{2}z}\label{Complex exponent span}\\
\int d^{2}\boldsymbol{q}\phi_{lp}^{*}\left(\boldsymbol{q}\right)\phi_{mn}\left(\boldsymbol{q}\right) & = & \delta_{lm}\delta_{pn}\label{Orthonormality}\\
\sum_{l,p}\phi_{lp}^{*}\left(\boldsymbol{q}\right)\phi_{lp}\left(\boldsymbol{q}'\right) & = & \delta\left(\boldsymbol{q}-\boldsymbol{q}'\right)\label{Closure}
\end{eqnarray}

\end{subequations}

\noindent where $\boldsymbol{q}$ is the transverse momentum and $\phi_{lp}\left(\boldsymbol{q}\right)$
is the LG spatial modes Fourier transform taken at $z=0$. The full
Hamiltonian is given by ${\cal H}=H_{0}+H_{\mu\phi}$, where $H_{0}=H_{\phi}+H_{\mu}$
is the noninteracting Hamiltonian of the radiation $\left(\phi\right)$ and
matter $\left(\mu\right)$. With these notations, the
interaction Hamiltonian can be recast in the form,

\begin{align}
H_{\mu\phi} & =\sum_{\boldsymbol{k}\boldsymbol{q}}\sigma_{\boldsymbol{k}-\boldsymbol{q}}^{\dagger}a_{\boldsymbol{k}}^{\dagger}a_{\boldsymbol{q}}+\sigma_{\boldsymbol{k}-\boldsymbol{q}}a_{\boldsymbol{k}}a_{\boldsymbol{q}}^{\dagger},\label{H_int 2nd quant}
\end{align}

\noindent where the highly oscillating terms corresponding to $a_{\boldsymbol{k}}a_{\boldsymbol{q}}$
and $a_{\boldsymbol{k}}^{\dagger}a_{\boldsymbol{q}}^{\dagger}$ are
neglected. The matter Hamiltonian is given by $H_{\mu}=\sum_{\alpha,i}\epsilon_{i,\alpha}c_{i,\alpha}^{\dagger}c_{i,\alpha}$,
where $c_{i,\alpha}\left(c_{i,\alpha}^{\dagger}\right)$ is the bosonic
annihilation (creation) operator with the canonical commutation relations
$\left[c_{i,\alpha},c_{j,\beta}^{\dagger}\right]=\delta_{\alpha\beta}\delta_{i,j}$.
The vibrational modes labeled $i$, and scatters by $\alpha$. The charge-density operator reads $\sigma_{\boldsymbol{k}}=\sum_{\alpha=1}^{N}f_{\alpha}\left(\boldsymbol{k}\right)c_{i,\alpha}^{\dagger}c_{j,\alpha},$
where $f_{\alpha}\left(\boldsymbol{k}\right)=e^{i\boldsymbol{k}\cdot\boldsymbol{r}_{\alpha}}w\left(\boldsymbol{k}\right)$,
$w\left(\boldsymbol{k}\right)$ are the localized molecular orbital
and $\boldsymbol{r}_{\alpha}$ are scatterers positions. We then calculate
the effective photon-photon interaction using the Schrieffer-Wolff
transformation {[}see appendix (B.1) for detailed derivation{]}, follows
by a transformation of the momentum representation into superposition
the SBs. 

\emph{The SB representation.} Using the above definitions
for the Bogoliubov transformation from momentum space to the Schmidt
representation $a_{\boldsymbol{k}}=\sum_{n}\phi_{n}\left(\boldsymbol{k}\right)a_{n},$
where $a_{n}$ is the Schmidt boson annihilation operator and $n$
is a shorthand notation for the two quantum numbers $l,p$ introduced
in the expression for the vector potential. The free Hamiltonian reads
assuming a single longitudinal mode at $\omega_{0}$ we get $H_{\phi}=\omega_{0}\sum_{n}a_{n}^{\dagger}a_{n}.$
The off-diagonal contributions to Eq.$\left(\text{\ref{Effective Hamiltoian}}\right)$ - namely, the hopping terms - are given by (see appendix for detailed derivation), 

\begin{subequations}

\begin{align}
\hat{t}_{coh} & =\sum_{nm}\theta_{nm}^{coh}a_{n}^{\dagger}a_{m}+h.c.,\\
\hat{t}_{inc} & =\sum_{nm}\theta_{nm}^{inc}\left(\Delta_{i}\right)a_{n}^{\dagger}a_{m}+\theta_{nm}^{inc*}\left(-\Delta_{i}\right)a_{m}^{\dagger}a_{n},
\end{align}

\end{subequations}

\noindent with the hopping coefficients are,

\begin{subequations}

\begin{align}
\theta_{nm}^{coh} & =2\sum_{\alpha\neq\beta}\sum_{\boldsymbol{k},\boldsymbol{q},\boldsymbol{s}}e^{-i\boldsymbol{\left(k-\boldsymbol{q}\right)}\cdot\boldsymbol{r}_{\alpha}+i\left(\boldsymbol{s}-\boldsymbol{q}\right)\cdot\boldsymbol{r}_{\beta}}\label{eq:exact coh}\\
\times & \frac{\phi_{n}^{*}\left(\boldsymbol{k}\right)w_{\alpha}^{*}\left(\boldsymbol{k}-\boldsymbol{q}\right)w_{\beta}\left(\boldsymbol{s}-\boldsymbol{q}\right)\phi_{m}\left(\boldsymbol{s}\right)}{q^{2}-k^{2}},\nonumber 
\end{align}

\begin{align}
\theta_{nm}^{inc} & \left(\Delta_{i}\right)=2\sum_{\alpha}\sum_{\boldsymbol{k},\boldsymbol{q},\boldsymbol{s}}e^{-i\left(\boldsymbol{k}-\boldsymbol{s}\right)\cdot\boldsymbol{r}_{\alpha}}\label{eq:Exact inc}\\
\times & \frac{\phi_{n}^{*}\left(\boldsymbol{k}\right)w_{\alpha}^{*}\left(\boldsymbol{k}-\boldsymbol{q'}\right)w_{\alpha}\left(\boldsymbol{q}-\boldsymbol{q}'\right)\phi_{m}\left(\boldsymbol{s}\right)}{q'^{2}-\left(k^{2}+\Delta_{i}\right)},\nonumber 
\end{align}

\end{subequations}

\noindent and $\Delta_{i}=2\omega_{0}\epsilon_{ig}$. Assuming a slowly
varying orbital in momentum domain (point particle limit), significantly
simplifies Eqs.$\left(\text{\ref{eq:exact coh}},\text{\ref{eq:Exact inc}}\right)$.
When the inter-particle distance is much smaller than the transverse
wavector $r_{\alpha\beta}\ll\lambda_{\perp}$, the coherent hopping
{[}Eq.$\left(\text{\ref{eq:exact coh}}\right)${]} is granted an intuitive
form, carrying spatial contributions due to the phase-difference of
scattered modes. Introducing a cutoff frequency $\Lambda$ such that
$\nicefrac{\left|r_{\beta\alpha}\right|}{\Lambda}\le\left|r_{\beta\alpha}\right|k\ll1$
and $q>\nicefrac{1}{\Lambda}$ we obtain,

\begin{subequations}
\begin{align}
\theta_{nm}^{coh} & =g^{coh}\sum_{\alpha\neq\beta}\psi_{n}^{*}\left(\boldsymbol{r}_{\alpha}\right)\psi_{m}\left(\boldsymbol{r}_{\beta}\right),\label{eq:coherent hopping term}
\end{align}

\begin{equation}
\theta_{nk}^{inc}=g^{inc}\sum_{\alpha}\psi_{n}^{*}\left(r_{\alpha}\right)\psi_{k}\left(r_{\alpha}\right).\label{eq:Incoherent hopping term}
\end{equation}

\end{subequations}

\noindent Note that these approximations are not essential yet grant
important intuition; see appendix B for the exact expressions.
The over-all SB hopping coefficient is given by $\Theta_{nk}=\omega_{0}\left(\delta_{nk}-\theta_{nk}^{coh}-\theta_{nk}^{inc}\right)$.

The interaction term in Eq.$\left(\text{\ref{Effective Hamiltoian}}\right)$ can be displayed in a form that separates the
geometrical factor from the basis dependent one $U_{nmls}=-\sum_{l^{'}s^{'}}S_{ll^{'}ss^{'}}V_{nml^{'}s^{'}}$.
Here $V_{nlkm}$ is the basis dependent four-mode \emph{scattering
potential},

\begin{align}
V_{nlkm} & =\sum_{i}\int d^{2}\boldsymbol{k}d^{2}\boldsymbol{q}\:\phi_{n}^{*}\left(\boldsymbol{k}\right)\phi_{k}\left(\boldsymbol{k}\right)\label{4-modes Scattering potential}\\
\times & \left[\frac{1}{k^{2}-q^{2}-\Delta_{i}}-\frac{1}{k^{2}-q^{2}+\Delta_{i}}\right]\phi_{l}\left(\boldsymbol{q}\right)\phi_{m}^{*}\left(\boldsymbol{q}\right),\nonumber 
\end{align}

and the geometric structural tensor, $S_{ll^{'}ss^{'}}=\sum_{\alpha}f_{ll'}^{\alpha*}f_{ss'}^{\alpha},$
using concatenation of basis $\leftrightarrow$ geometry transformation
$f_{nm}^{\alpha}=\psi_{n}^{*}\left(r_{\alpha}\right)\psi_{m}\left(r_{\alpha}\right)$
- isolating basis dependent properties from the geometric characteristics.
The effective hamiltonian is finally given by Eq.$\left(\text{\ref{Effective Hamiltoian}}\right)$.

\begin{figure}[h]
\begin{centering}
(a)\includegraphics[bb=110bp 270bp 450bp 540bp,clip,scale=0.67]{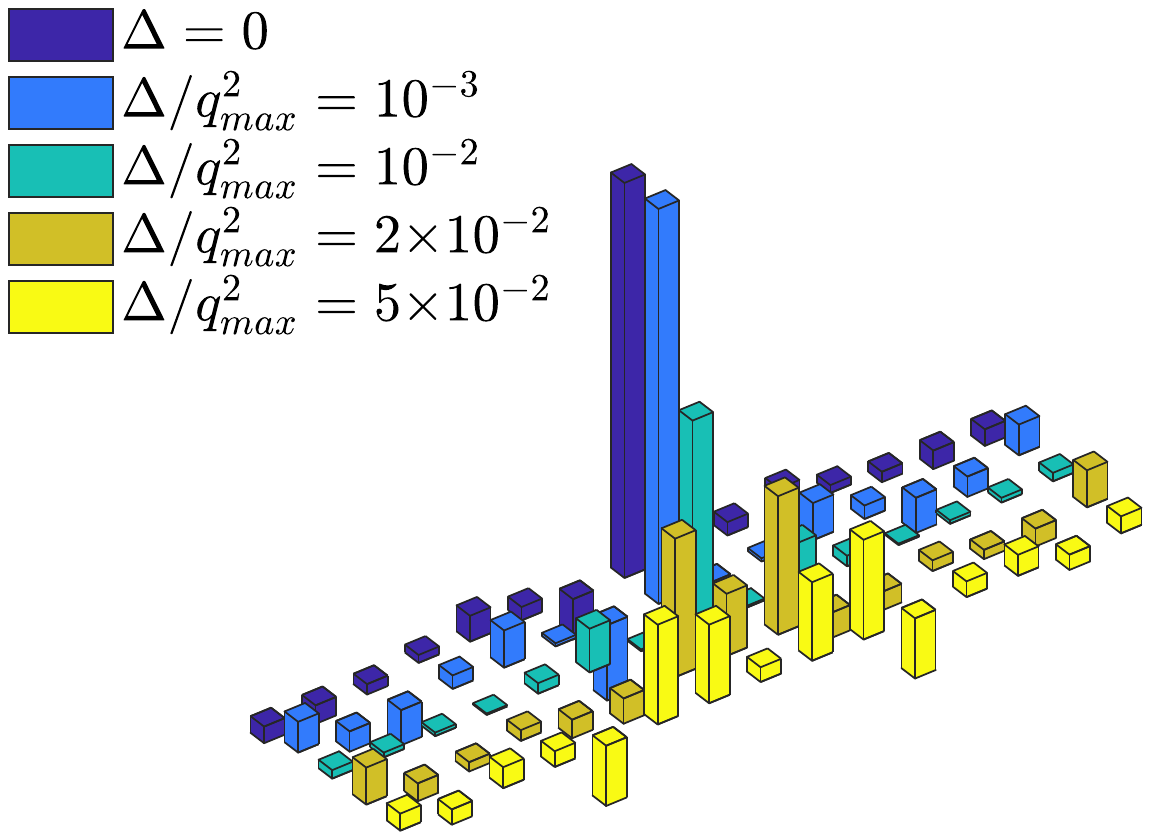}
\par\end{centering}
\begin{centering}
(b)\includegraphics[bb=140bp 260bp 490bp 540bp,clip,scale=0.3]{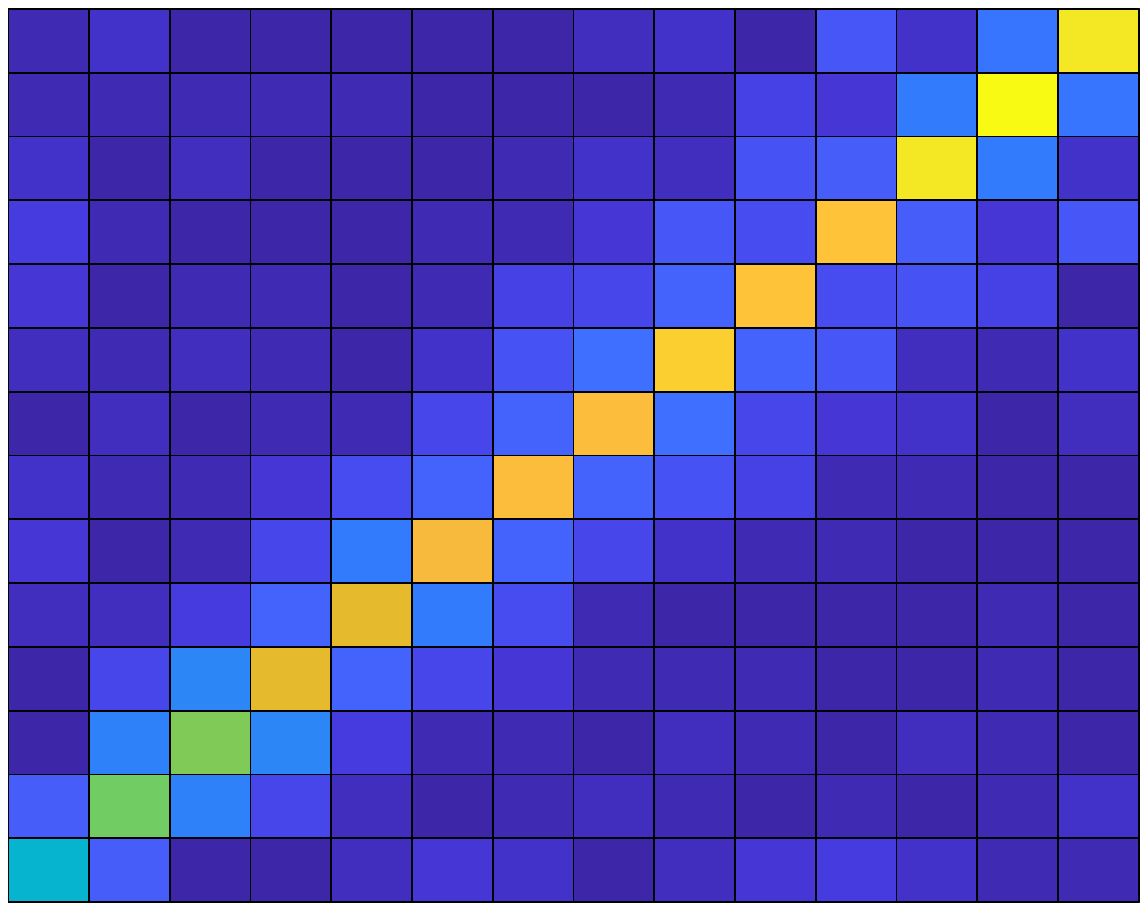}(c)\includegraphics[bb=140bp 260bp 490bp 540bp,clip,scale=0.3]{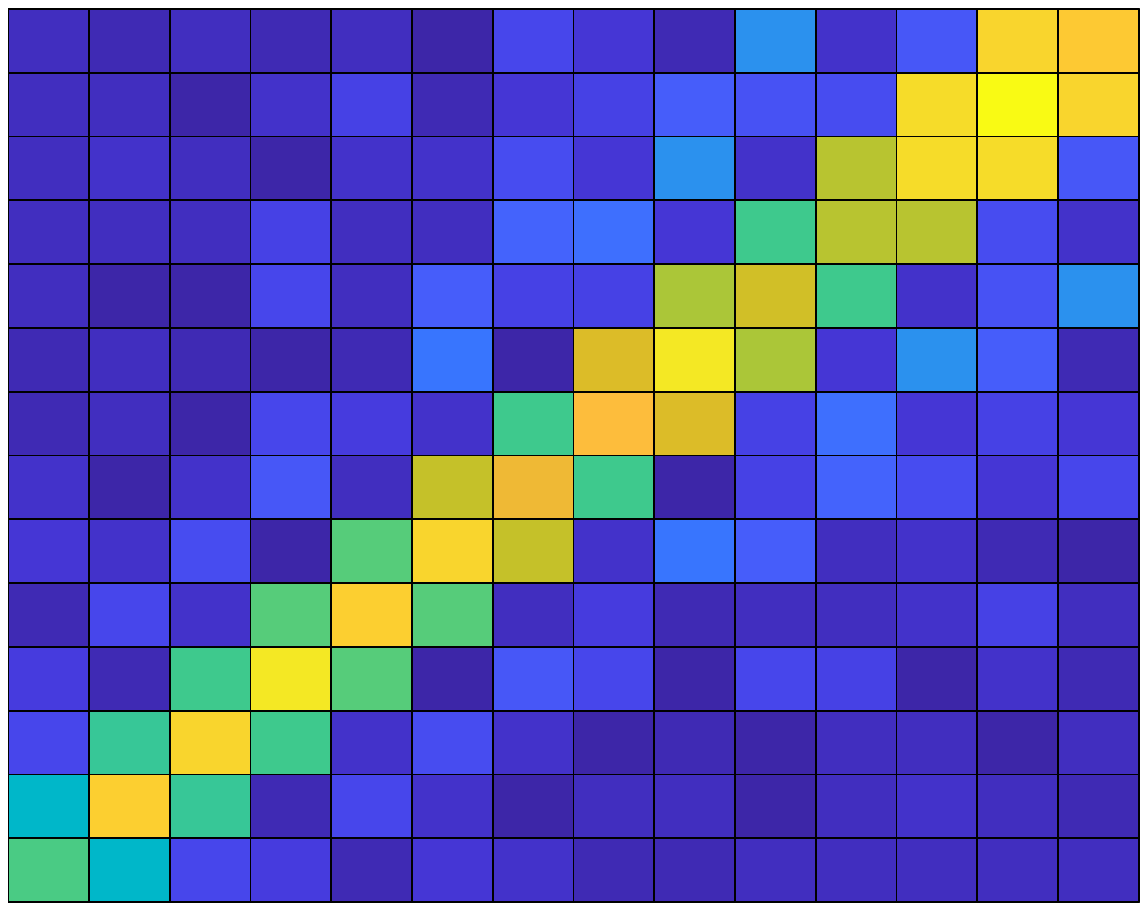}
\par\end{centering}
\begin{centering}
(d)\includegraphics[bb=140bp 260bp 490bp 540bp,clip,scale=0.3]{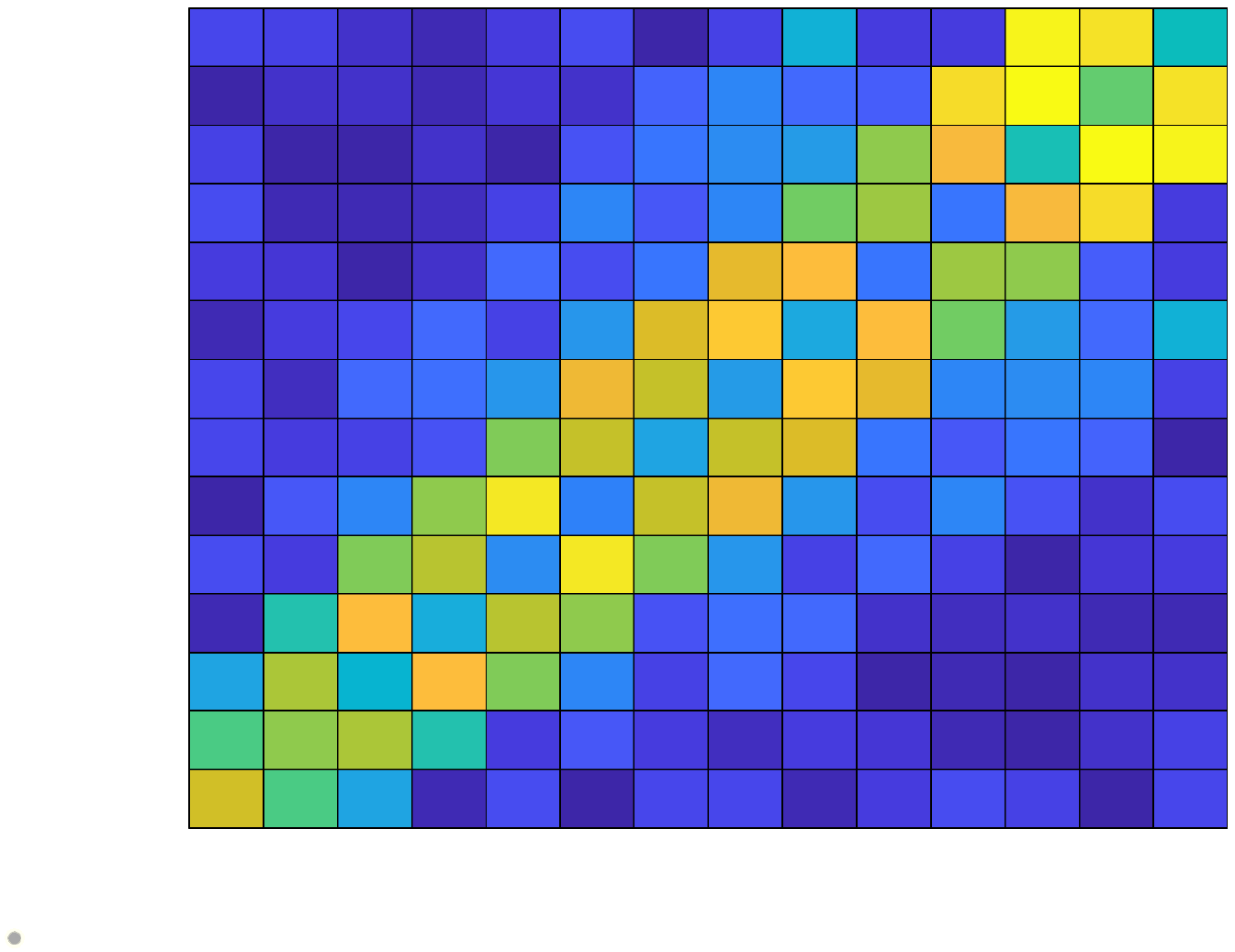}(e)\includegraphics[bb=140bp 260bp 490bp 540bp,clip,scale=0.3]{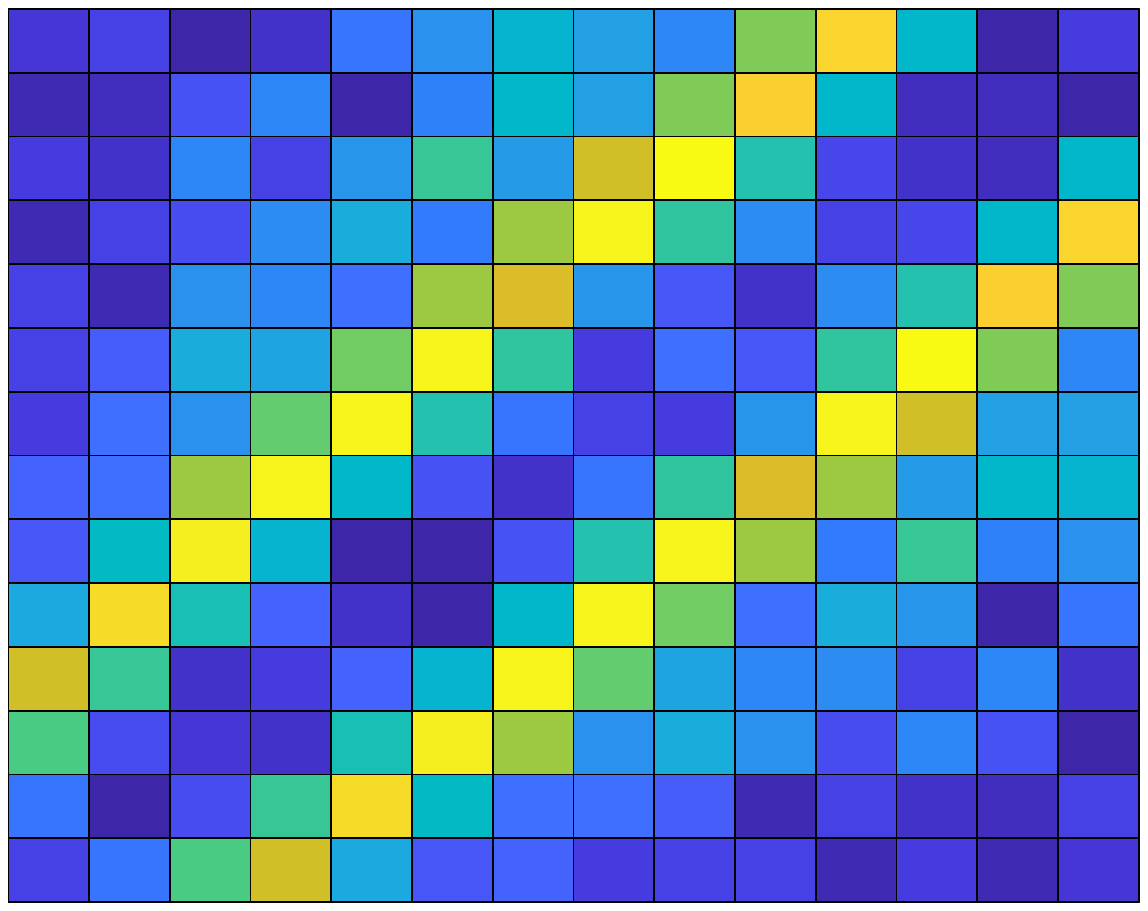}
\par\end{centering}
\caption{\emph{The scattering potential.} $V_{nmkl}$ in Eq.$\left(\text{\ref{4-modes Scattering potential}}\right)$
calculated in the LG basis. Each panel corresponds to different $\Delta$
value, and panel (a) captures a summary of the diagonal crossection
from the top-left to bottom-right of panels b:e. Panels b:e capture
the inter-mode coupling {[}relations between the two middle dimensions
$\left(V_{mk}=tr_{nl}V_{nmkl}\right)${]}, of the effective SBs scattering
potential for the selected values $\nicefrac{\Delta}{\boldsymbol{q}_{max}^{2}}=\left(0.001,0.01,0.02,0.05\right)$.
The calculated relations between the first two $\left(V_{nm}=tr_{kl}V_{nmkl}\right)$
and last dimensions $\left(V_{kl}=tr_{nm}V_{nmkl}\right)$ are indistinguishable
from panel (b) for all values of $\Delta$ and resemble a Kronecker
delta distribution. \label{LG scattering-potential}}
\end{figure}
\emph{The scattering potential in the Laguerre-Gauss basis.} The target
effective interaction to be simulated is manipulated and designed
using three main ingredients. (i) One is purely geometric and defined
by the arrangement of charges, (ii) the internal structure of
each charge $\left(\Delta_{i}\right)$ and (iii) the choice
of basis. While the geometric component appears in the expressions
for both, the interaction and hopping terms, the charge spectral structure
contributes to the scattering potential of Eq.$\left(\text{\ref{4-modes Scattering potential}}\right)$
alone. We now study the effects of $\Delta$ on the interaction which
is basis dependent yet geometry independent. The scattering potential
$V_{nklm}$ involves coupling between four modes. Eq.$\left(\text{\ref{4-modes Scattering potential}}\right)$
can be further simplified  using the LG basis {[}see Eq.$\left(35B\right)$
of the supplementary material{]}. The structure of the scattering
potential as a function of $\Delta$ for a two level system is depicted
in Fig. $\left(\text{\ref{LG scattering-potential}}\right)$. For
all values of $\Delta,$the relation between the first and last two
modes {[}$n,k$ and $l,m$ of Eq.$\left(\text{\ref{Effective Hamiltoian}}\right)${]}
resembles a Kronecker delta . It is given by the partial traces $V_{nm}=tr_{kl}V_{nmkl}\approx\delta_{nm}$
and $V_{kl}=tr_{nm}V_{nmkl}\approx\delta_{kl}$ and depicted in Fig.$\left(\text{\ref{LG scattering-potential}}a\right)$.
This property simplifies the effective Hamiltonian to,

\begin{equation}
H_{eff}^{LG}=\sum_{n,k}\Theta_{nk}a_{n}^{\dagger}a_{k}-\sum_{nk}\mathcal{U}_{nk}\hat{n}_{n}\hat{n}_{k},
\end{equation}

\noindent where $\hat{n}_{k}=a_{k}^{\dagger}a_{k}$ is the number
operator, and ${\cal U}_{nk}=\sum_{lm}U_{nlkm}\delta_{nl}\delta_{km}$.
The effective potential between the first and last two modes of Eq.$\left(\text{\ref{Effective Hamiltoian}}\right)$
spreads to neighboring modes with increasing values of $\nicefrac{\Delta}{\boldsymbol{q}_{max}^{2}}$
where $\boldsymbol{q}_{max}$ is the cutoff wavector in the numerical
calculation. This behavior is summarized in Fig.$\left(\text{\ref{LG scattering-potential}}a\right)$,
and demonstrated separately for the selected values in Fig.$\left(\text{\ref{LG scattering-potential}}b:e\right)$.
At large values of $\nicefrac{\Delta}{\boldsymbol{q}_{max}^{2}}$
the scattering occurs between more distant modes, corresponding to
energy exchange with the matter. Extension of this result to a system
of charges composed of a more complex internal structure, is given
by a straightforward summation, resulting in longer-range interaction.\\

\begin{figure}[h]
\begin{centering}
\includegraphics[bb=0bp 0bp 841bp 576bp,clip,scale=0.2]{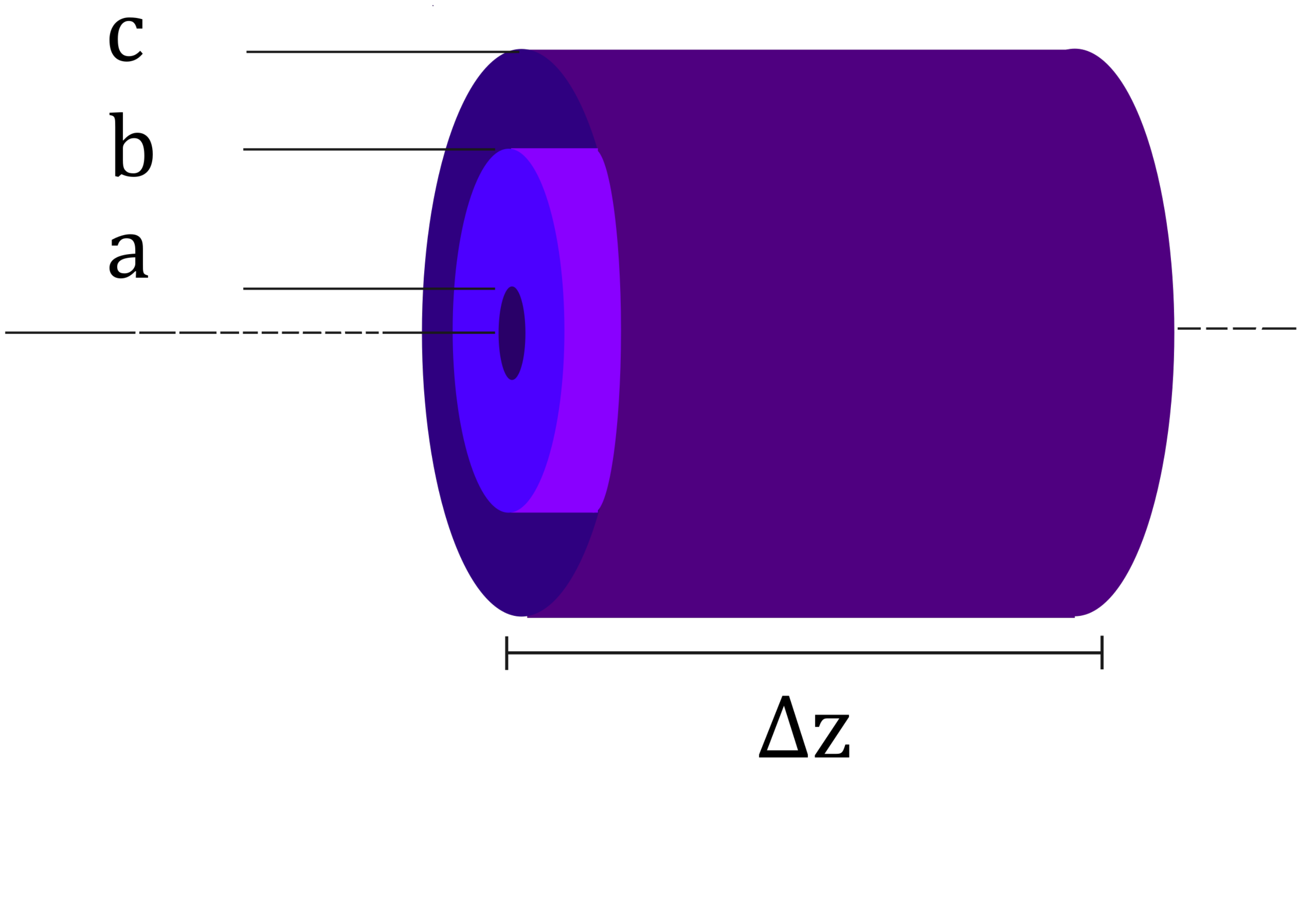}
\par\end{centering}
\caption{\emph{Uniformly distributed hollow-cylinder}. Inner and outer radius
$a$ and $b$ respectively, $c$ is chosen such that 95\% of the power
of the radial mode for which $n=25$ is contained within. $\Delta z$
is the region in which the interaction and hopping occur, corresponding
to the interaction time interval $\tau$. \label{fig:Illustrative-example Uniform cylinder sketch}}
\end{figure}

\emph{Illustrative example of the effective Hamiltonian.} We derive the effective Hamiltonian for molecules in cylindrical
architecture, in which controlled hopping can confine the dynamics to
a restricted subspace. We consider a uniform distribution of
molecules filling a hollow-cylinder (UC) of inner radius $a$ and
outer radius $b$ as shown in Fig.$\left(\text{\ref{fig:Illustrative-example Uniform cylinder sketch}}\right)$.
The boundary radius $c$ is defined such that 95\% of the power of
the incident radial mode for which $n=25$ is contained within the
calculation range of the numerical simulation (the chosen cutoff mode).
 
\begin{figure}[th]
\begin{centering}
(a)\includegraphics[bb=150bp 270bp 500bp 540bp,clip,scale=0.35]{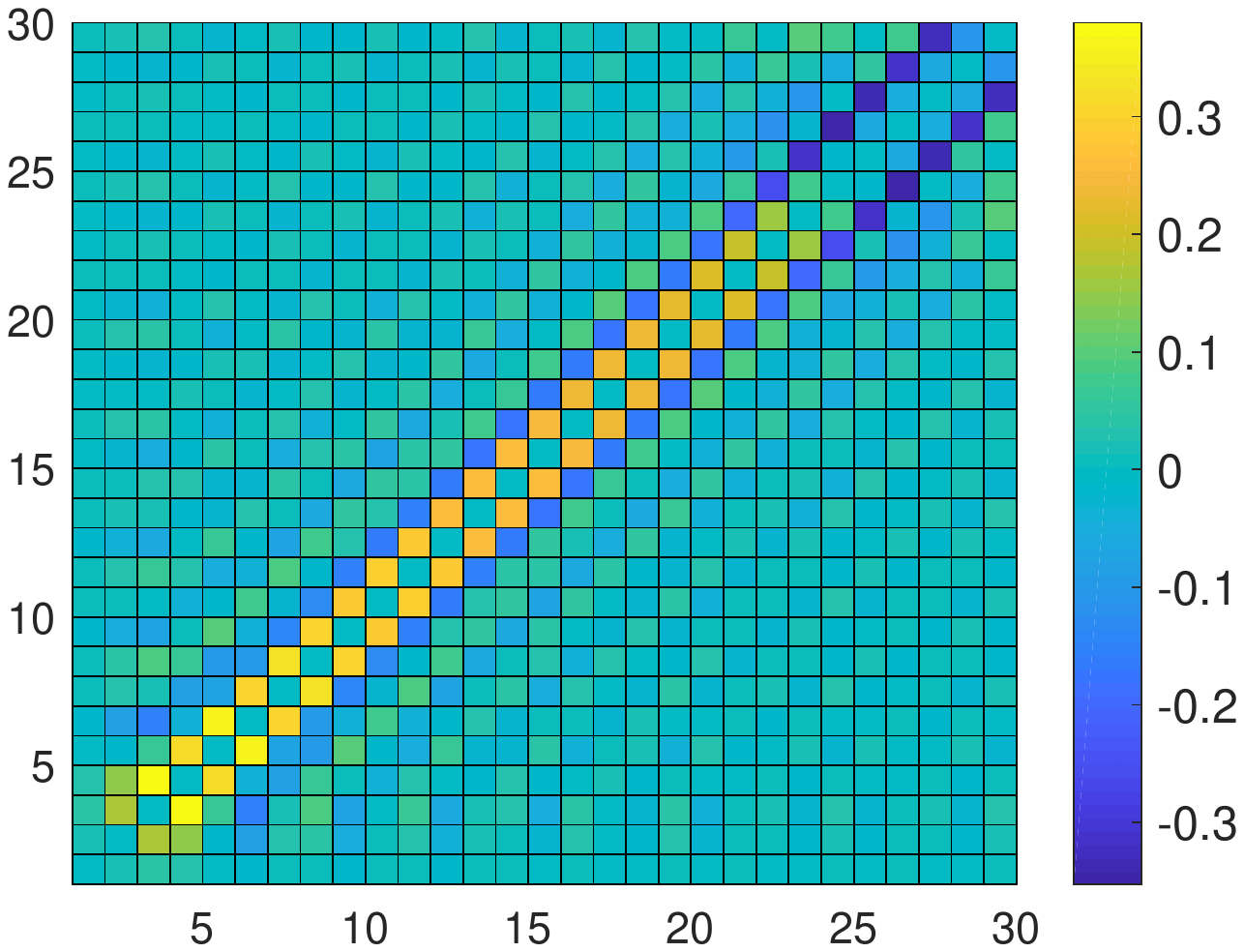}(b)\includegraphics[bb=150bp 270bp 500bp 540bp,clip,scale=0.35]{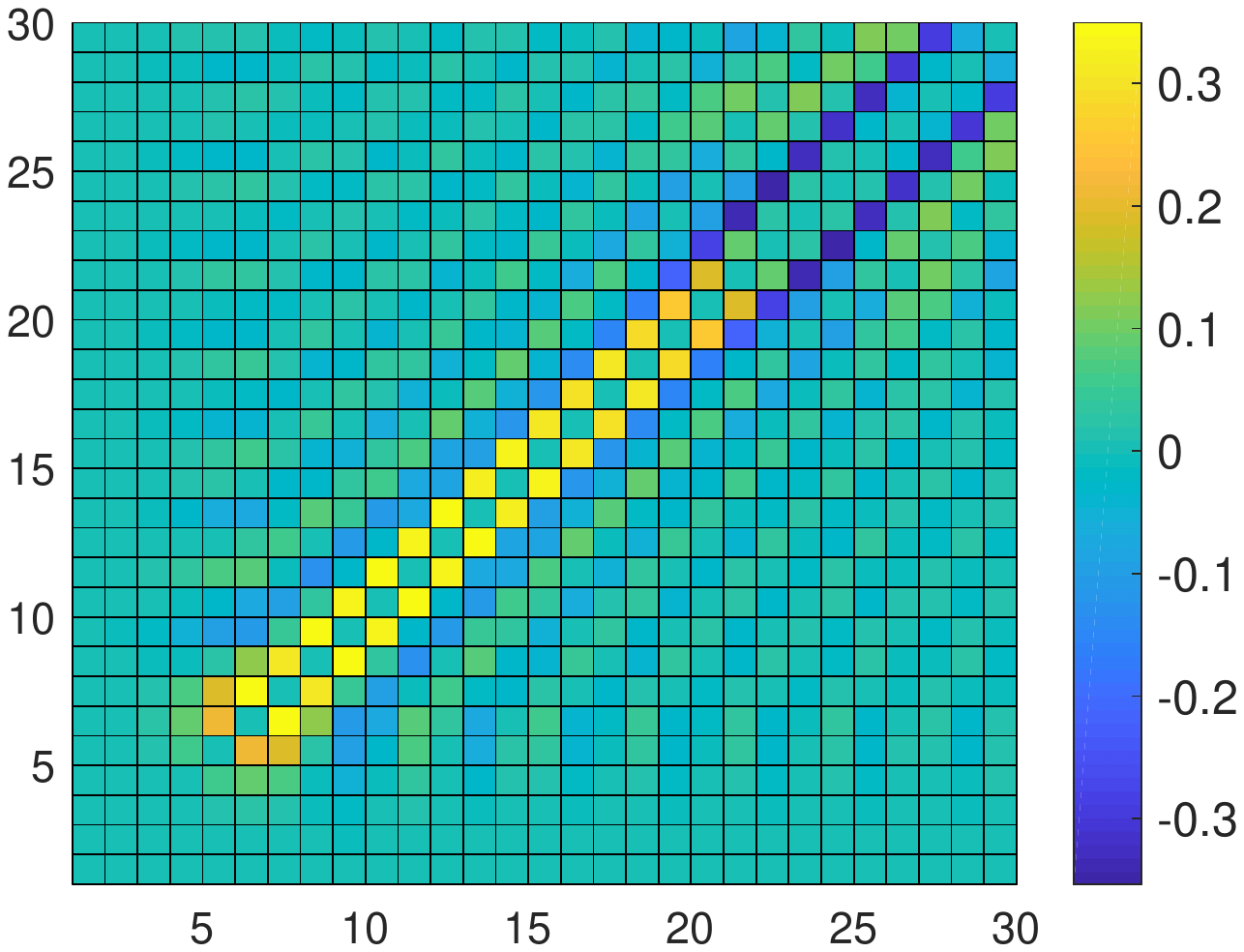}
\par\end{centering}
\begin{centering}
\par\end{centering}
\begin{centering}
(c)\includegraphics[bb=150bp 270bp 500bp 540bp,clip,scale=0.35]{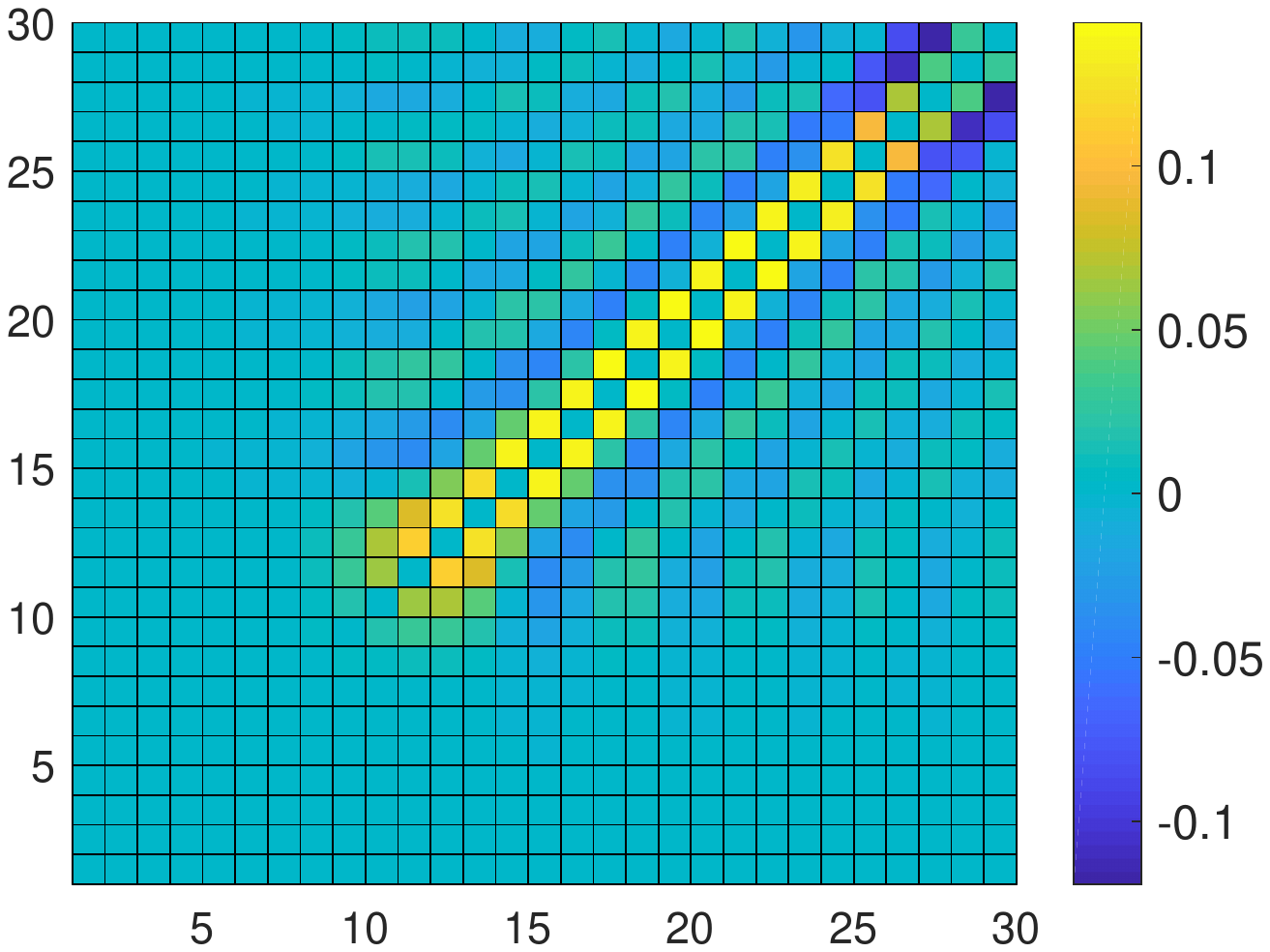}(d)\includegraphics[bb=150bp 270bp 500bp 540bp,clip,scale=0.35]{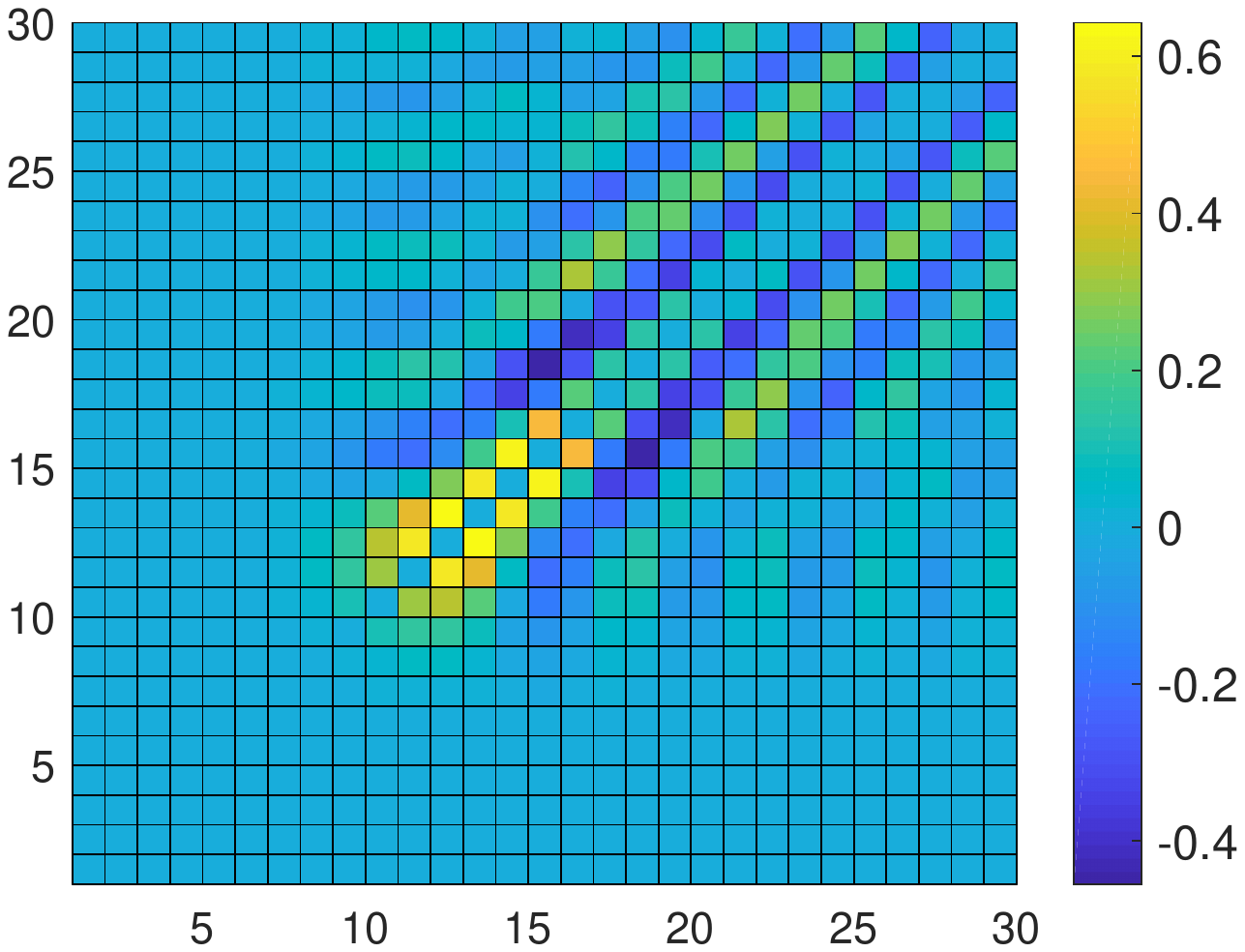}
\par\end{centering}
\caption{\emph{Geometrically controlled hopping confinement}. (a-d) The geometric
hopping factors $\theta_{nk}^{inc}$ presented in Eq.$\left(\text{\ref{eq:Incoherent hopping term}}\right)$
are displayed for the above geometry. Panels a-d computed with the
corresponding dimensionless distance from the origin $\left(\nicefrac{a}{c},\nicefrac{b}{c}\right)$:
$\left[\text{a}\left(0.1,0.9\right),\text{b}\left(0.2,0.8\right),\text{c}\left(0.4,1\right),\text{d}\left(0.4,0.6\right)\right]$.
Positive contributions are energetically favorable. \label{fig:Hopping term from Uniform cylinder}}
\end{figure}
In this case the geometric coefficient of the hopping terms in Eqs.$\left(\text{\ref{eq:coherent hopping term}},\text{\ref{eq:Incoherent hopping term}}\right)$
as well as the interaction can be used to confine the dynamics in
a controlled subspace. By varying $a$ and $b$, the hopping range
depicted in Figs.$\left(\text{\ref{fig:Hopping term from Uniform cylinder}}\right)$
a:d can be controlled. Fig.$\left(\text{\ref{fig:Hopping term from Uniform cylinder}}\right)$
along with Eqs.$\left(\text{\ref{eq:coherent hopping term}},\text{\ref{eq:Incoherent hopping term}}\right)$
show that hopping within the set of modes corresponding to positive
geometric factor, is energetically favorable. These positive contributions
are surrounded by negative ones which are costly energetically, resulting
in effective confinement. In this setup the single-molecule field
scattering presented in Eq.$\left(\text{\ref{eq:Incoherent hopping term}}\right)$
dominates the hopping dynamics and the coherent hopping term of Eq.$\left(\text{\ref{eq:coherent hopping term}}\right)$
is suppressed. This is verified by the vanishing structure factor
in a disordered lattice, or equivalently from the closure relations
of the LG basis combined with orthogonality {[}Eqs.$\left(\text{\ref{Closure}},\text{\ref{Orthonormality}}\right)${]}. Using Eq.$\left(\text{\ref{Complex exponent span}}\right)$ one can estimate that for $\left|\boldsymbol{q}_{max}\right|=10^{-p}k_{0}$ and $\Delta z=10^{l}\lambda_{0}$ the modal attenuation factor is $\exp\left(-2\pi10^{l-2p}\right)$ which  yields $\approx 94\%$ of the incoming photon flux at the output for $ l=p=2 $.    
This geometry simulates the dynamics of the Hamiltonian $H_{eff}^{LG}=H_{UC}$,\\
\begin{equation}
H_{UC}=\sum_{\left\langle n,k\right\rangle \in{\cal D}_{h}}\Theta_{nk}a_{n}^{\dagger}a_{k}-\sum_{n,k\in{\cal D}_{\Delta}}\mathcal{U}_{nk}\hat{n}_{n}\hat{n}_{k},
\end{equation}
where $\left\langle n,k\right\rangle $ stands for nearest neighbors.
${\cal D}_{h}$ is a domain determined by the geometry in which the
hopping occurs, as shown in Fig.$\left(\text{\ref{fig:Hopping term from Uniform cylinder}}\right)$.
${\cal D}_{\Delta}$ is the domain set by $\Delta$ for which several
illustrations are depicted in Fig.$\left(\text{\ref{LG scattering-potential}}\right)$.
\\

\emph{Discussion.} We have developed a geometric SPC, shaping photon-photon interactions via geometric design of the coupling between spatial modes, using the setup depicted in Fig.$\left(\text{\ref{fig:Setup}}\right)$. Quantum dynamics of interacting bosons described by the Hamiltonian of Eq.$\left(\text{\ref{Effective Hamiltoian}}\right)$ can be simulated and directly measured using the above ingredients.

The dynamics of the SBs constrained by the Eq.$\left(\text{\ref{Effective Hamiltoian}}\right)$
 is controlled by the following three main quantities. The geometric distribution
of molecules, their internal structure and the choice of spatial basis. The dynamics induced by the geometric SPC depicted in  Fig.$\left(\text{\ref{fig:Illustrative-example Uniform cylinder sketch}}\right)$  can be restricted to a finite set of modes as demonstrated in Fig.$\left(\text{\ref{fig:Hopping term from Uniform cylinder}}\right)$. This offers a \emph{purpose-computing platform}
to a class of  problems with exponential complexity. There is a growing interest
in purpose machines, built for the solution of a specific task, e.g.
coherent Ising machines \citep{Inagaki603,McMahon614}. These structures
are designed to solve efficiently Ising models on graphs with programmable
connectivity. Their usefulness stems from the well known mapping between
the Ising model ground state search problem, and combinatorial optimization
problems in polynomial time \citep{Barahona_1982} (both NP hard).
The proposed setup is also applicable as a quantum (light) state-preparation
technique, as well as multi-photon gate in a photonic quantum processor. 

In molecular systems the number of vibrational modes $N_{vib}$ is
proportional to the number of atoms $N_{a}$ according to $N_{vib}=3N_{a}-6$.
The number of electronic states corresponds to the number of electrons
$N_{e}$. Good candidates would be systems containing few vibrational
modes while large number of electrons that potentially provide strong
coupling of the vibrational modes with applied electromagnetic field.
Short wavelength tabletop X-ray sources that couple off-resonantly
between the vibrational modes provide intriguing possibility for source
realization \cite{Rocca:16}. Longer wavelength sources for which sophisticated measurement
techniques are more mature may be possible although the coupling
between the modes may take more complicated forms.

Due to the structure of the LG modes, for low number of ordered scatterers,
sign-flipping anti-ferromagnetic coupling has been observed that requires
further characterization. Finding the molecular distribution and basis
emulating the desired dynamics provides a topic for future study as
well as coupling to electronic states rather than the vibrational. 
For this purpose, another degree of the geometric properties could
be considered, the local charge distribution of a each scatterer presented
in Eqs.$\left(\text{\ref{eq:exact coh},\ref{eq:Exact inc}}\right)$.

\textbf{\emph{Acknowledgments.}} The support of the Chemical Sciences,
Geosciences, and Biosciences Division, Office of Basic Energy Sciences,
Office of Science, U.S. Department of Energy is gratefully acknowledged.
S.M was supported by Award DE-FG02-04ER15571. S.A fellowship was supported
by the National Science Foundation (Grant No. CHE-1663822).

\bibliography{Geo_SPC_bib}

\begin{thebibliography}{29}%
\makeatletter
\providecommand \@ifxundefined [1]{%
 \@ifx{#1\undefined}
}%
\providecommand \@ifnum [1]{%
 \ifnum #1\expandafter \@firstoftwo
 \else \expandafter \@secondoftwo
 \fi
}%
\providecommand \@ifx [1]{%
 \ifx #1\expandafter \@firstoftwo
 \else \expandafter \@secondoftwo
 \fi
}%
\providecommand \natexlab [1]{#1}%
\providecommand \enquote  [1]{``#1''}%
\providecommand \bibnamefont  [1]{#1}%
\providecommand \bibfnamefont [1]{#1}%
\providecommand \citenamefont [1]{#1}%
\providecommand \href@noop [0]{\@secondoftwo}%
\providecommand \href [0]{\begingroup \@sanitize@url \@href}%
\providecommand \@href[1]{\@@startlink{#1}\@@href}%
\providecommand \@@href[1]{\endgroup#1\@@endlink}%
\providecommand \@sanitize@url [0]{\catcode `\\12\catcode `\$12\catcode
  `\&12\catcode `\#12\catcode `\^12\catcode `\_12\catcode `\%12\relax}%
\providecommand \@@startlink[1]{}%
\providecommand \@@endlink[0]{}%
\providecommand \url  [0]{\begingroup\@sanitize@url \@url }%
\providecommand \@url [1]{\endgroup\@href {#1}{\urlprefix }}%
\providecommand \urlprefix  [0]{URL }%
\providecommand \Eprint [0]{\href }%
\providecommand \doibase [0]{https://doi.org/}%
\providecommand \selectlanguage [0]{\@gobble}%
\providecommand \bibinfo  [0]{\@secondoftwo}%
\providecommand \bibfield  [0]{\@secondoftwo}%
\providecommand \translation [1]{[#1]}%
\providecommand \BibitemOpen [0]{}%
\providecommand \bibitemStop [0]{}%
\providecommand \bibitemNoStop [0]{.\EOS\space}%
\providecommand \EOS [0]{\spacefactor3000\relax}%
\providecommand \BibitemShut  [1]{\csname bibitem#1\endcsname}%
\let\auto@bib@innerbib\@empty
\bibitem [{\citenamefont {Nielsen}\ and\ \citenamefont
  {Chuang}(2010)}]{nielsen_chuang_2010}%
  \BibitemOpen
  \bibfield  {author} {\bibinfo {author} {\bibfnamefont {M.~A.}\ \bibnamefont
  {Nielsen}}\ and\ \bibinfo {author} {\bibfnamefont {I.~L.}\ \bibnamefont
  {Chuang}},\ }\href {https://doi.org/10.1017/CBO9780511976667} {\emph
  {\bibinfo {title} {{Quantum Computation and Quantum Information: 10th
  Anniversary Edition}}}}\ (\bibinfo  {publisher} {Cambridge University
  Press},\ \bibinfo {year} {2010})\BibitemShut {NoStop}%
\bibitem [{Note1()}]{Note1}%
  \BibitemOpen
  \bibinfo {note} {Assuming voltage levels are assigned to bit states, having
  '0' represented by zero voltage.}\BibitemShut {Stop}%
\bibitem [{\citenamefont {Massoud~Salehi}\ and\ \citenamefont
  {Proakis}(2007)}]{massoud2007digital}%
  \BibitemOpen
  \bibfield  {author} {\bibinfo {author} {\bibfnamefont {P.}~\bibnamefont
  {Massoud~Salehi}}\ and\ \bibinfo {author} {\bibfnamefont {J.}~\bibnamefont
  {Proakis}},\ }\href {https://books.google.com/books?id=HroiQAAACAAJ} {\emph
  {\bibinfo {title} {Digital Communications}}}\ (\bibinfo  {publisher}
  {McGraw-Hill Education},\ \bibinfo {year} {2007})\BibitemShut {NoStop}%
\bibitem [{\citenamefont {Shor}(1995)}]{PhysRevA.52.R2493}%
  \BibitemOpen
  \bibfield  {author} {\bibinfo {author} {\bibfnamefont {P.~W.}\ \bibnamefont
  {Shor}},\ }\bibfield  {title} {\bibinfo {title} {{Scheme for reducing
  decoherence in quantum computer memory}},\ }\href
  {https://doi.org/10.1103/PhysRevA.52.R2493} {\bibfield  {journal} {\bibinfo
  {journal} {Phys. Rev. A}\ }\textbf {\bibinfo {volume} {52}},\ \bibinfo
  {pages} {R2493} (\bibinfo {year} {1995})}\BibitemShut {NoStop}%
\bibitem [{\citenamefont {Gottesman}\ \emph {et~al.}(2001)\citenamefont
  {Gottesman}, \citenamefont {Kitaev},\ and\ \citenamefont
  {Preskill}}]{PhysRevA.64.012310}%
  \BibitemOpen
  \bibfield  {author} {\bibinfo {author} {\bibfnamefont {D.}~\bibnamefont
  {Gottesman}}, \bibinfo {author} {\bibfnamefont {A.}~\bibnamefont {Kitaev}},\
  and\ \bibinfo {author} {\bibfnamefont {J.}~\bibnamefont {Preskill}},\
  }\bibfield  {title} {\bibinfo {title} {{Encoding a qubit in an oscillator}},\
  }\href {https://doi.org/10.1103/PhysRevA.64.012310} {\bibfield  {journal}
  {\bibinfo  {journal} {Phys. Rev. A}\ }\textbf {\bibinfo {volume} {64}},\
  \bibinfo {pages} {12310} (\bibinfo {year} {2001})}\BibitemShut {NoStop}%
\bibitem [{\citenamefont {Schindler}\ \emph {et~al.}(2011)\citenamefont
  {Schindler}, \citenamefont {Barreiro}, \citenamefont {Monz}, \citenamefont
  {Nebendahl}, \citenamefont {Nigg}, \citenamefont {Chwalla}, \citenamefont
  {Hennrich},\ and\ \citenamefont {Blatt}}]{Schindler1059}%
  \BibitemOpen
  \bibfield  {author} {\bibinfo {author} {\bibfnamefont {P.}~\bibnamefont
  {Schindler}}, \bibinfo {author} {\bibfnamefont {J.~T.}\ \bibnamefont
  {Barreiro}}, \bibinfo {author} {\bibfnamefont {T.}~\bibnamefont {Monz}},
  \bibinfo {author} {\bibfnamefont {V.}~\bibnamefont {Nebendahl}}, \bibinfo
  {author} {\bibfnamefont {D.}~\bibnamefont {Nigg}}, \bibinfo {author}
  {\bibfnamefont {M.}~\bibnamefont {Chwalla}}, \bibinfo {author} {\bibfnamefont
  {M.}~\bibnamefont {Hennrich}},\ and\ \bibinfo {author} {\bibfnamefont
  {R.}~\bibnamefont {Blatt}},\ }\bibfield  {title} {\bibinfo {title}
  {{Experimental Repetitive Quantum Error Correction}},\ }\href
  {https://doi.org/10.1126/science.1203329} {\bibfield  {journal} {\bibinfo
  {journal} {Science}\ }\textbf {\bibinfo {volume} {332}},\ \bibinfo {pages}
  {1059} (\bibinfo {year} {2011})}\BibitemShut {NoStop}%
\bibitem [{\citenamefont {Cirac}\ and\ \citenamefont
  {Zoller}(1995)}]{PhysRevLett.74.4091}%
  \BibitemOpen
  \bibfield  {author} {\bibinfo {author} {\bibfnamefont {J.~I.}\ \bibnamefont
  {Cirac}}\ and\ \bibinfo {author} {\bibfnamefont {P.}~\bibnamefont {Zoller}},\
  }\bibfield  {title} {\bibinfo {title} {{Quantum Computations with Cold
  Trapped Ions}},\ }\href {https://doi.org/10.1103/PhysRevLett.74.4091}
  {\bibfield  {journal} {\bibinfo  {journal} {Phys. Rev. Lett.}\ }\textbf
  {\bibinfo {volume} {74}},\ \bibinfo {pages} {4091} (\bibinfo {year}
  {1995})}\BibitemShut {NoStop}%
\bibitem [{\citenamefont {Jaksch}\ \emph {et~al.}(1998)\citenamefont {Jaksch},
  \citenamefont {Bruder}, \citenamefont {Cirac}, \citenamefont {Gardiner},\
  and\ \citenamefont {Zoller}}]{PhysRevLett.81.3108}%
  \BibitemOpen
  \bibfield  {author} {\bibinfo {author} {\bibfnamefont {D.}~\bibnamefont
  {Jaksch}}, \bibinfo {author} {\bibfnamefont {C.}~\bibnamefont {Bruder}},
  \bibinfo {author} {\bibfnamefont {J.~I.}\ \bibnamefont {Cirac}}, \bibinfo
  {author} {\bibfnamefont {C.~W.}\ \bibnamefont {Gardiner}},\ and\ \bibinfo
  {author} {\bibfnamefont {P.}~\bibnamefont {Zoller}},\ }\bibfield  {title}
  {\bibinfo {title} {{Cold Bosonic Atoms in Optical Lattices}},\ }\href
  {https://doi.org/10.1103/PhysRevLett.81.3108} {\bibfield  {journal} {\bibinfo
   {journal} {Phys. Rev. Lett.}\ }\textbf {\bibinfo {volume} {81}},\ \bibinfo
  {pages} {3108} (\bibinfo {year} {1998})}\BibitemShut {NoStop}%
\bibitem [{\citenamefont {Monroe}(2002)}]{Monroe2002}%
  \BibitemOpen
  \bibfield  {author} {\bibinfo {author} {\bibfnamefont {C.}~\bibnamefont
  {Monroe}},\ }\bibfield  {title} {\bibinfo {title} {{Quantum information
  processing with atoms and photons}},\ }\href {https://doi.org/10.1038/416238a
  http://10.0.4.14/416238a} {\bibfield  {journal} {\bibinfo  {journal}
  {Nature}\ }\textbf {\bibinfo {volume} {416}},\ \bibinfo {pages} {238}
  (\bibinfo {year} {2002})}\BibitemShut {NoStop}%
\bibitem [{\citenamefont {Kim}\ \emph {et~al.}(2010)\citenamefont {Kim},
  \citenamefont {Chang}, \citenamefont {Korenblit}, \citenamefont {Islam},
  \citenamefont {Edwards}, \citenamefont {Freericks}, \citenamefont {Lin},
  \citenamefont {Duan},\ and\ \citenamefont {Monroe}}]{Kim2010}%
  \BibitemOpen
  \bibfield  {author} {\bibinfo {author} {\bibfnamefont {K.}~\bibnamefont
  {Kim}}, \bibinfo {author} {\bibfnamefont {M.-S.}\ \bibnamefont {Chang}},
  \bibinfo {author} {\bibfnamefont {S.}~\bibnamefont {Korenblit}}, \bibinfo
  {author} {\bibfnamefont {R.}~\bibnamefont {Islam}}, \bibinfo {author}
  {\bibfnamefont {E.~E.}\ \bibnamefont {Edwards}}, \bibinfo {author}
  {\bibfnamefont {J.~K.}\ \bibnamefont {Freericks}}, \bibinfo {author}
  {\bibfnamefont {G.-D.}\ \bibnamefont {Lin}}, \bibinfo {author} {\bibfnamefont
  {L.-M.}\ \bibnamefont {Duan}},\ and\ \bibinfo {author} {\bibfnamefont
  {C.}~\bibnamefont {Monroe}},\ }\bibfield  {title} {\bibinfo {title} {{Quantum
  simulation of frustrated Ising spins with trapped ions}},\ }\href
  {https://doi.org/10.1038/nature09071 http://10.0.4.14/nature09071
  https://www.nature.com/articles/nature09071{\#}supplementary-information}
  {\bibfield  {journal} {\bibinfo  {journal} {Nature}\ }\textbf {\bibinfo
  {volume} {465}},\ \bibinfo {pages} {590} (\bibinfo {year}
  {2010})}\BibitemShut {NoStop}%
\bibitem [{\citenamefont {Schreiber}\ \emph {et~al.}(2015)\citenamefont
  {Schreiber}, \citenamefont {Hodgman}, \citenamefont {Bordia}, \citenamefont
  {L{\"{u}}schen}, \citenamefont {Fischer}, \citenamefont {Vosk}, \citenamefont
  {Altman}, \citenamefont {Schneider},\ and\ \citenamefont
  {Bloch}}]{Schreiber842}%
  \BibitemOpen
  \bibfield  {author} {\bibinfo {author} {\bibfnamefont {M.}~\bibnamefont
  {Schreiber}}, \bibinfo {author} {\bibfnamefont {S.~S.}\ \bibnamefont
  {Hodgman}}, \bibinfo {author} {\bibfnamefont {P.}~\bibnamefont {Bordia}},
  \bibinfo {author} {\bibfnamefont {H.~P.}\ \bibnamefont {L{\"{u}}schen}},
  \bibinfo {author} {\bibfnamefont {M.~H.}\ \bibnamefont {Fischer}}, \bibinfo
  {author} {\bibfnamefont {R.}~\bibnamefont {Vosk}}, \bibinfo {author}
  {\bibfnamefont {E.}~\bibnamefont {Altman}}, \bibinfo {author} {\bibfnamefont
  {U.}~\bibnamefont {Schneider}},\ and\ \bibinfo {author} {\bibfnamefont
  {I.}~\bibnamefont {Bloch}},\ }\bibfield  {title} {\bibinfo {title}
  {{Observation of many-body localization of interacting fermions in a
  quasirandom optical lattice}},\ }\href
  {https://doi.org/10.1126/science.aaa7432} {\bibfield  {journal} {\bibinfo
  {journal} {Science}\ }\textbf {\bibinfo {volume} {349}},\ \bibinfo {pages}
  {842} (\bibinfo {year} {2015})}\BibitemShut {NoStop}%
\bibitem [{\citenamefont {Smith}\ \emph {et~al.}(2016)\citenamefont {Smith},
  \citenamefont {Lee}, \citenamefont {Richerme}, \citenamefont {Neyenhuis},
  \citenamefont {Hess}, \citenamefont {Hauke}, \citenamefont {Heyl},
  \citenamefont {Huse},\ and\ \citenamefont {Monroe}}]{Smith2016}%
  \BibitemOpen
  \bibfield  {author} {\bibinfo {author} {\bibfnamefont {J.}~\bibnamefont
  {Smith}}, \bibinfo {author} {\bibfnamefont {A.}~\bibnamefont {Lee}}, \bibinfo
  {author} {\bibfnamefont {P.}~\bibnamefont {Richerme}}, \bibinfo {author}
  {\bibfnamefont {B.}~\bibnamefont {Neyenhuis}}, \bibinfo {author}
  {\bibfnamefont {P.~W.}\ \bibnamefont {Hess}}, \bibinfo {author}
  {\bibfnamefont {P.}~\bibnamefont {Hauke}}, \bibinfo {author} {\bibfnamefont
  {M.}~\bibnamefont {Heyl}}, \bibinfo {author} {\bibfnamefont {D.~A.}\
  \bibnamefont {Huse}},\ and\ \bibinfo {author} {\bibfnamefont
  {C.}~\bibnamefont {Monroe}},\ }\bibfield  {title} {\bibinfo {title}
  {{Many-body localization in a quantum simulator with programmable random
  disorder}},\ }\href {https://doi.org/10.1038/nphys3783
  http://10.0.4.14/nphys3783
  https://www.nature.com/articles/nphys3783{\#}supplementary-information}
  {\bibfield  {journal} {\bibinfo  {journal} {Nature Physics}\ }\textbf
  {\bibinfo {volume} {12}},\ \bibinfo {pages} {907} (\bibinfo {year}
  {2016})}\BibitemShut {NoStop}%
\bibitem [{\citenamefont {Zohar}\ \emph {et~al.}(2015)\citenamefont {Zohar},
  \citenamefont {Cirac},\ and\ \citenamefont {Reznik}}]{Zohar_2015}%
  \BibitemOpen
  \bibfield  {author} {\bibinfo {author} {\bibfnamefont {E.}~\bibnamefont
  {Zohar}}, \bibinfo {author} {\bibfnamefont {J.~I.}\ \bibnamefont {Cirac}},\
  and\ \bibinfo {author} {\bibfnamefont {B.}~\bibnamefont {Reznik}},\
  }\bibfield  {title} {\bibinfo {title} {{Quantum simulations of lattice gauge
  theories using ultracold atoms in optical lattices}},\ }\href
  {https://doi.org/10.1088/0034-4885/79/1/014401} {\bibfield  {journal}
  {\bibinfo  {journal} {Reports on Progress in Physics}\ }\textbf {\bibinfo
  {volume} {79}},\ \bibinfo {pages} {14401} (\bibinfo {year}
  {2015})}\BibitemShut {NoStop}%
\bibitem [{\citenamefont {Rico}\ \emph {et~al.}(2018)\citenamefont {Rico},
  \citenamefont {Dalmonte}, \citenamefont {Zoller}, \citenamefont {Banerjee},
  \citenamefont {Bogli}, \citenamefont {Stebler},\ and\ \citenamefont
  {Wiese}}]{RICO2018}%
  \BibitemOpen
  \bibfield  {author} {\bibinfo {author} {\bibfnamefont {E.}~\bibnamefont
  {Rico}}, \bibinfo {author} {\bibfnamefont {M.}~\bibnamefont {Dalmonte}},
  \bibinfo {author} {\bibfnamefont {P.}~\bibnamefont {Zoller}}, \bibinfo
  {author} {\bibfnamefont {D.}~\bibnamefont {Banerjee}}, \bibinfo {author}
  {\bibfnamefont {M.}~\bibnamefont {Bogli}}, \bibinfo {author} {\bibfnamefont
  {P.}~\bibnamefont {Stebler}},\ and\ \bibinfo {author} {\bibfnamefont {U.-J.}\
  \bibnamefont {Wiese}},\ }\bibfield  {title} {\bibinfo {title} {So(3) "nuclear
  physics" with ultracold gases},\ }\href
  {https://doi.org/https://doi.org/10.1016/j.aop.2018.03.020} {\bibfield
  {journal} {\bibinfo  {journal} {Annals of Physics}\ }\textbf {\bibinfo
  {volume} {393}},\ \bibinfo {pages} {466 } (\bibinfo {year}
  {2018})}\BibitemShut {NoStop}%
\bibitem [{\citenamefont {Muschik}\ \emph {et~al.}(2017)\citenamefont
  {Muschik}, \citenamefont {Heyl}, \citenamefont {Martinez}, \citenamefont
  {Monz}, \citenamefont {Schindler}, \citenamefont {Vogell}, \citenamefont
  {Dalmonte}, \citenamefont {Hauke}, \citenamefont {Blatt},\ and\ \citenamefont
  {Zoller}}]{Muschik_2017}%
  \BibitemOpen
  \bibfield  {author} {\bibinfo {author} {\bibfnamefont {C.}~\bibnamefont
  {Muschik}}, \bibinfo {author} {\bibfnamefont {M.}~\bibnamefont {Heyl}},
  \bibinfo {author} {\bibfnamefont {E.}~\bibnamefont {Martinez}}, \bibinfo
  {author} {\bibfnamefont {T.}~\bibnamefont {Monz}}, \bibinfo {author}
  {\bibfnamefont {P.}~\bibnamefont {Schindler}}, \bibinfo {author}
  {\bibfnamefont {B.}~\bibnamefont {Vogell}}, \bibinfo {author} {\bibfnamefont
  {M.}~\bibnamefont {Dalmonte}}, \bibinfo {author} {\bibfnamefont
  {P.}~\bibnamefont {Hauke}}, \bibinfo {author} {\bibfnamefont
  {R.}~\bibnamefont {Blatt}},\ and\ \bibinfo {author} {\bibfnamefont
  {P.}~\bibnamefont {Zoller}},\ }\bibfield  {title} {\bibinfo {title} {U(1)
  wilson lattice gauge theories in digital quantum simulators},\ }\href
  {https://doi.org/10.1088/1367-2630/aa89ab} {\bibfield  {journal} {\bibinfo
  {journal} {New Journal of Physics}\ }\textbf {\bibinfo {volume} {19}},\
  \bibinfo {pages} {103020} (\bibinfo {year} {2017})}\BibitemShut {NoStop}%
\bibitem [{Note2()}]{Note2}%
  \BibitemOpen
  \bibinfo {note} {In linear order of the paraxiality parameter $\vartheta $
  there are \protect \emph {no losses} in the longitudinal modes. Losses in the
  transverse modes are governed solely by the geometry. Higher orders exhibit
  losses and demonstrated in the illustration below.}\BibitemShut {Stop}%
\bibitem [{\citenamefont {Asban}\ \emph {et~al.}(2019)\citenamefont {Asban},
  \citenamefont {Dorfman},\ and\ \citenamefont {Mukamel}}]{Asban2019}%
  \BibitemOpen
  \bibfield  {author} {\bibinfo {author} {\bibfnamefont {S.}~\bibnamefont
  {Asban}}, \bibinfo {author} {\bibfnamefont {K.~E.}\ \bibnamefont {Dorfman}},\
  and\ \bibinfo {author} {\bibfnamefont {S.}~\bibnamefont {Mukamel}},\
  }\bibfield  {title} {\bibinfo {title} {Quantum phase-sensitive diffraction
  and imaging using entangled photons},\ }\href
  {https://doi.org/10.1073/pnas.1904839116} {\bibfield  {journal} {\bibinfo
  {journal} {Proceedings of the National Academy of Sciences}\ }\textbf
  {\bibinfo {volume} {116}},\ \bibinfo {pages} {11673} (\bibinfo {year}
  {2019})},\ \Eprint
  {https://arxiv.org/abs/https://www.pnas.org/content/116/24/11673.full.pdf}
  {https://www.pnas.org/content/116/24/11673.full.pdf} \BibitemShut {NoStop}%
\bibitem [{\citenamefont {Fickler}\ \emph {et~al.}(2016)\citenamefont
  {Fickler}, \citenamefont {Campbell}, \citenamefont {Buchler}, \citenamefont
  {Lam},\ and\ \citenamefont {Zeilinger}}]{Fickler2016}%
  \BibitemOpen
  \bibfield  {author} {\bibinfo {author} {\bibfnamefont {R.}~\bibnamefont
  {Fickler}}, \bibinfo {author} {\bibfnamefont {G.}~\bibnamefont {Campbell}},
  \bibinfo {author} {\bibfnamefont {B.}~\bibnamefont {Buchler}}, \bibinfo
  {author} {\bibfnamefont {P.~K.}\ \bibnamefont {Lam}},\ and\ \bibinfo {author}
  {\bibfnamefont {A.}~\bibnamefont {Zeilinger}},\ }\bibfield  {title} {\bibinfo
  {title} {{Quantum entanglement of angular momentum states with quantum
  numbers up to 10,010}},\ }\href {https://doi.org/10.1073/pnas.1616889113}
  {\bibfield  {journal} {\bibinfo  {journal} {Proceedings of the National
  Academy of Sciences}\ }\textbf {\bibinfo {volume} {113}},\ \bibinfo {pages}
  {13642} (\bibinfo {year} {2016})},\ \Eprint
  {https://arxiv.org/abs/1607.00922} {arXiv:1607.00922} \BibitemShut {NoStop}%
\bibitem [{\citenamefont {Fickler}\ \emph {et~al.}(2012)\citenamefont
  {Fickler}, \citenamefont {Lapkiewicz}, \citenamefont {Plick}, \citenamefont
  {Krenn}, \citenamefont {Schaeff}, \citenamefont {Ramelow},\ and\
  \citenamefont {Zeilinger}}]{Fickler640}%
  \BibitemOpen
  \bibfield  {author} {\bibinfo {author} {\bibfnamefont {R.}~\bibnamefont
  {Fickler}}, \bibinfo {author} {\bibfnamefont {R.}~\bibnamefont {Lapkiewicz}},
  \bibinfo {author} {\bibfnamefont {W.~N.}\ \bibnamefont {Plick}}, \bibinfo
  {author} {\bibfnamefont {M.}~\bibnamefont {Krenn}}, \bibinfo {author}
  {\bibfnamefont {C.}~\bibnamefont {Schaeff}}, \bibinfo {author} {\bibfnamefont
  {S.}~\bibnamefont {Ramelow}},\ and\ \bibinfo {author} {\bibfnamefont
  {A.}~\bibnamefont {Zeilinger}},\ }\bibfield  {title} {\bibinfo {title}
  {{Quantum Entanglement of High Angular Momenta}},\ }\href
  {https://doi.org/10.1126/science.1227193} {\bibfield  {journal} {\bibinfo
  {journal} {Science}\ }\textbf {\bibinfo {volume} {338}},\ \bibinfo {pages}
  {640} (\bibinfo {year} {2012})}\BibitemShut {NoStop}%
\bibitem [{\citenamefont {Krenn}\ \emph {et~al.}(2013)\citenamefont {Krenn},
  \citenamefont {Fickler}, \citenamefont {Huber}, \citenamefont {Lapkiewicz},
  \citenamefont {Plick}, \citenamefont {Ramelow},\ and\ \citenamefont
  {Zeilinger}}]{PhysRevA.87.012326}%
  \BibitemOpen
  \bibfield  {author} {\bibinfo {author} {\bibfnamefont {M.}~\bibnamefont
  {Krenn}}, \bibinfo {author} {\bibfnamefont {R.}~\bibnamefont {Fickler}},
  \bibinfo {author} {\bibfnamefont {M.}~\bibnamefont {Huber}}, \bibinfo
  {author} {\bibfnamefont {R.}~\bibnamefont {Lapkiewicz}}, \bibinfo {author}
  {\bibfnamefont {W.}~\bibnamefont {Plick}}, \bibinfo {author} {\bibfnamefont
  {S.}~\bibnamefont {Ramelow}},\ and\ \bibinfo {author} {\bibfnamefont
  {A.}~\bibnamefont {Zeilinger}},\ }\bibfield  {title} {\bibinfo {title}
  {{Entangled singularity patterns of photons in Ince-Gauss modes}},\ }\href
  {https://doi.org/10.1103/PhysRevA.87.012326} {\bibfield  {journal} {\bibinfo
  {journal} {Phys. Rev. A}\ }\textbf {\bibinfo {volume} {87}},\ \bibinfo
  {pages} {12326} (\bibinfo {year} {2013})}\BibitemShut {NoStop}%
\bibitem [{\citenamefont {Bavaresco}\ \emph {et~al.}(2018)\citenamefont
  {Bavaresco}, \citenamefont {{Herrera Valencia}}, \citenamefont
  {Kl{\"{o}}ckl}, \citenamefont {Pivoluska}, \citenamefont {Erker},
  \citenamefont {Friis}, \citenamefont {Malik},\ and\ \citenamefont
  {Huber}}]{Bavaresco2018}%
  \BibitemOpen
  \bibfield  {author} {\bibinfo {author} {\bibfnamefont {J.}~\bibnamefont
  {Bavaresco}}, \bibinfo {author} {\bibfnamefont {N.}~\bibnamefont {{Herrera
  Valencia}}}, \bibinfo {author} {\bibfnamefont {C.}~\bibnamefont
  {Kl{\"{o}}ckl}}, \bibinfo {author} {\bibfnamefont {M.}~\bibnamefont
  {Pivoluska}}, \bibinfo {author} {\bibfnamefont {P.}~\bibnamefont {Erker}},
  \bibinfo {author} {\bibfnamefont {N.}~\bibnamefont {Friis}}, \bibinfo
  {author} {\bibfnamefont {M.}~\bibnamefont {Malik}},\ and\ \bibinfo {author}
  {\bibfnamefont {M.}~\bibnamefont {Huber}},\ }\bibfield  {title} {\bibinfo
  {title} {{Measurements in two bases are sufficient for certifying
  high-dimensional entanglement}},\ }\href
  {https://doi.org/10.1038/s41567-018-0203-z} {\bibfield  {journal} {\bibinfo
  {journal} {Nature Physics}\ }\textbf {\bibinfo {volume} {14}},\ \bibinfo
  {pages} {1032} (\bibinfo {year} {2018})}\BibitemShut {NoStop}%
\bibitem [{\citenamefont {Straupe}\ \emph {et~al.}(2011)\citenamefont
  {Straupe}, \citenamefont {Ivanov}, \citenamefont {Kalinkin}, \citenamefont
  {Bobrov},\ and\ \citenamefont {Kulik}}]{Straupe2011}%
  \BibitemOpen
  \bibfield  {author} {\bibinfo {author} {\bibfnamefont {S.~S.}\ \bibnamefont
  {Straupe}}, \bibinfo {author} {\bibfnamefont {D.~P.}\ \bibnamefont {Ivanov}},
  \bibinfo {author} {\bibfnamefont {A.~A.}\ \bibnamefont {Kalinkin}}, \bibinfo
  {author} {\bibfnamefont {I.~B.}\ \bibnamefont {Bobrov}},\ and\ \bibinfo
  {author} {\bibfnamefont {S.~P.}\ \bibnamefont {Kulik}},\ }\bibfield  {title}
  {\bibinfo {title} {Angular schmidt modes in spontaneous parametric
  down-conversion},\ }\href {https://doi.org/10.1103/PhysRevA.83.060302}
  {\bibfield  {journal} {\bibinfo  {journal} {Phys. Rev. A}\ }\textbf {\bibinfo
  {volume} {83}},\ \bibinfo {pages} {060302(R)} (\bibinfo {year}
  {2011})}\BibitemShut {NoStop}%
\bibitem [{Note3()}]{Note3}%
  \BibitemOpen
  \bibinfo {note} {Using entangled photon pair such that the as SB and RSB
  ideally results in background free signal yet harder to produce.}\BibitemShut
  {Stop}%
\bibitem [{\citenamefont {Aiello}\ and\ \citenamefont
  {Woerdman}(2005)}]{PhysRevA.72.060101}%
  \BibitemOpen
  \bibfield  {author} {\bibinfo {author} {\bibfnamefont {A.}~\bibnamefont
  {Aiello}}\ and\ \bibinfo {author} {\bibfnamefont {J.~P.}\ \bibnamefont
  {Woerdman}},\ }\bibfield  {title} {\bibinfo {title} {Exact quantization of a
  paraxial electromagnetic field},\ }\href
  {https://doi.org/10.1103/PhysRevA.72.060101} {\bibfield  {journal} {\bibinfo
  {journal} {Phys. Rev. A}\ }\textbf {\bibinfo {volume} {72}},\ \bibinfo
  {pages} {060101(R)} (\bibinfo {year} {2005})}\BibitemShut {NoStop}%
\bibitem [{\citenamefont {Calvo}\ \emph {et~al.}(2006)\citenamefont {Calvo},
  \citenamefont {Picon},\ and\ \citenamefont {Bagan}}]{Calvo2006}%
  \BibitemOpen
  \bibfield  {author} {\bibinfo {author} {\bibfnamefont {G.~F.}\ \bibnamefont
  {Calvo}}, \bibinfo {author} {\bibfnamefont {A.}~\bibnamefont {Picon}},\ and\
  \bibinfo {author} {\bibfnamefont {E.}~\bibnamefont {Bagan}},\ }\bibfield
  {title} {\bibinfo {title} {Quantum field theory of photons with orbital
  angular momentum},\ }\href {https://doi.org/10.1103/PhysRevA.73.013805}
  {\bibfield  {journal} {\bibinfo  {journal} {Phys. Rev. A}\ }\textbf {\bibinfo
  {volume} {73}},\ \bibinfo {pages} {013805} (\bibinfo {year}
  {2006})}\BibitemShut {NoStop}%
\bibitem [{\citenamefont {Inagaki}\ \emph {et~al.}(2016)\citenamefont
  {Inagaki}, \citenamefont {Haribara}, \citenamefont {Igarashi}, \citenamefont
  {Sonobe}, \citenamefont {Tamate}, \citenamefont {Honjo}, \citenamefont
  {Marandi}, \citenamefont {McMahon}, \citenamefont {Umeki}, \citenamefont
  {Enbutsu}, \citenamefont {Tadanaga}, \citenamefont {Takenouchi},
  \citenamefont {Aihara}, \citenamefont {Kawarabayashi}, \citenamefont {Inoue},
  \citenamefont {Utsunomiya},\ and\ \citenamefont {Takesue}}]{Inagaki603}%
  \BibitemOpen
  \bibfield  {author} {\bibinfo {author} {\bibfnamefont {T.}~\bibnamefont
  {Inagaki}}, \bibinfo {author} {\bibfnamefont {Y.}~\bibnamefont {Haribara}},
  \bibinfo {author} {\bibfnamefont {K.}~\bibnamefont {Igarashi}}, \bibinfo
  {author} {\bibfnamefont {T.}~\bibnamefont {Sonobe}}, \bibinfo {author}
  {\bibfnamefont {S.}~\bibnamefont {Tamate}}, \bibinfo {author} {\bibfnamefont
  {T.}~\bibnamefont {Honjo}}, \bibinfo {author} {\bibfnamefont
  {A.}~\bibnamefont {Marandi}}, \bibinfo {author} {\bibfnamefont {P.~L.}\
  \bibnamefont {McMahon}}, \bibinfo {author} {\bibfnamefont {T.}~\bibnamefont
  {Umeki}}, \bibinfo {author} {\bibfnamefont {K.}~\bibnamefont {Enbutsu}},
  \bibinfo {author} {\bibfnamefont {O.}~\bibnamefont {Tadanaga}}, \bibinfo
  {author} {\bibfnamefont {H.}~\bibnamefont {Takenouchi}}, \bibinfo {author}
  {\bibfnamefont {K.}~\bibnamefont {Aihara}}, \bibinfo {author} {\bibfnamefont
  {K.-i.}\ \bibnamefont {Kawarabayashi}}, \bibinfo {author} {\bibfnamefont
  {K.}~\bibnamefont {Inoue}}, \bibinfo {author} {\bibfnamefont
  {S.}~\bibnamefont {Utsunomiya}},\ and\ \bibinfo {author} {\bibfnamefont
  {H.}~\bibnamefont {Takesue}},\ }\bibfield  {title} {\bibinfo {title} {{A
  coherent Ising machine for 2000-node optimization problems}},\ }\href
  {https://doi.org/10.1126/science.aah4243} {\bibfield  {journal} {\bibinfo
  {journal} {Science}\ }\textbf {\bibinfo {volume} {354}},\ \bibinfo {pages}
  {603} (\bibinfo {year} {2016})}\BibitemShut {NoStop}%
\bibitem [{\citenamefont {McMahon}\ \emph {et~al.}(2016)\citenamefont
  {McMahon}, \citenamefont {Marandi}, \citenamefont {Haribara}, \citenamefont
  {Hamerly}, \citenamefont {Langrock}, \citenamefont {Tamate}, \citenamefont
  {Inagaki}, \citenamefont {Takesue}, \citenamefont {Utsunomiya}, \citenamefont
  {Aihara}, \citenamefont {Byer}, \citenamefont {Fejer}, \citenamefont
  {Mabuchi},\ and\ \citenamefont {Yamamoto}}]{McMahon614}%
  \BibitemOpen
  \bibfield  {author} {\bibinfo {author} {\bibfnamefont {P.~L.}\ \bibnamefont
  {McMahon}}, \bibinfo {author} {\bibfnamefont {A.}~\bibnamefont {Marandi}},
  \bibinfo {author} {\bibfnamefont {Y.}~\bibnamefont {Haribara}}, \bibinfo
  {author} {\bibfnamefont {R.}~\bibnamefont {Hamerly}}, \bibinfo {author}
  {\bibfnamefont {C.}~\bibnamefont {Langrock}}, \bibinfo {author}
  {\bibfnamefont {S.}~\bibnamefont {Tamate}}, \bibinfo {author} {\bibfnamefont
  {T.}~\bibnamefont {Inagaki}}, \bibinfo {author} {\bibfnamefont
  {H.}~\bibnamefont {Takesue}}, \bibinfo {author} {\bibfnamefont
  {S.}~\bibnamefont {Utsunomiya}}, \bibinfo {author} {\bibfnamefont
  {K.}~\bibnamefont {Aihara}}, \bibinfo {author} {\bibfnamefont {R.~L.}\
  \bibnamefont {Byer}}, \bibinfo {author} {\bibfnamefont {M.~M.}\ \bibnamefont
  {Fejer}}, \bibinfo {author} {\bibfnamefont {H.}~\bibnamefont {Mabuchi}},\
  and\ \bibinfo {author} {\bibfnamefont {Y.}~\bibnamefont {Yamamoto}},\
  }\bibfield  {title} {\bibinfo {title} {{A fully programmable 100-spin
  coherent Ising machine with all-to-all connections}},\ }\href
  {https://doi.org/10.1126/science.aah5178} {\bibfield  {journal} {\bibinfo
  {journal} {Science}\ }\textbf {\bibinfo {volume} {354}},\ \bibinfo {pages}
  {614} (\bibinfo {year} {2016})}\BibitemShut {NoStop}%
\bibitem [{\citenamefont {Barahona}(1982)}]{Barahona_1982}%
  \BibitemOpen
  \bibfield  {author} {\bibinfo {author} {\bibfnamefont {F.}~\bibnamefont
  {Barahona}},\ }\bibfield  {title} {\bibinfo {title} {{On the computational
  complexity of Ising spin glass models}},\ }\href
  {https://doi.org/10.1088/0305-4470/15/10/028} {\bibfield  {journal} {\bibinfo
   {journal} {Journal of Physics A: Mathematical and General}\ }\textbf
  {\bibinfo {volume} {15}},\ \bibinfo {pages} {3241} (\bibinfo {year}
  {1982})}\BibitemShut {NoStop}%
\bibitem [{\citenamefont {Rocca}(2016)}]{Rocca:16}%
  \BibitemOpen
  \bibfield  {author} {\bibinfo {author} {\bibfnamefont {J.~J.}\ \bibnamefont
  {Rocca}},\ }\bibfield  {title} {\bibinfo {title} {Table-top soft x-ray
  lasers},\ }in\ \href {https://doi.org/10.1364/CLEO_AT.2016.AM1K.1} {\emph
  {\bibinfo {booktitle} {Conference on Lasers and Electro-Optics}}}\ (\bibinfo
  {publisher} {Optical Society of America},\ \bibinfo {year} {2016})\ p.\
  \bibinfo {pages} {AM1K.1}\BibitemShut {NoStop}%
\end{thebibliography}%

\end{document}